\documentclass[aps,prl,floatfix,groupedaddress,twocolumn]{revtex4}
\usepackage{graphicx}
\usepackage[below]{placeins} 
\usepackage{textcomp}
\makeatletter
\renewcommand{\@fnsymbol}[1]{%
   \ifcase#1\or\textasteriskcentered\or\textsection\or\textdagger
   \or\textdaggerdbl\or\textparagraph\or\textbardbl
   \or\textasteriskcentered\textasteriskcentered
   \or\textdagger\textdagger\or\textdaggerdbl\textdaggerdbl
   \else\@ctrerr\fi}
\makeatother

\begin{document}
\graphicspath{{figs/}}
\DeclareGraphicsExtensions{.eps,.ps,.EPS,.PS}

\title{A high-resolution infrared spectroscopic investigation of the
   halogen atom-HCN entrance channel complexes solvated in superfluid
   helium droplets}

\author{Jeremy M. Merritt}%
\email{merritjm@unc.edu}%
\author{Jochen K\"upper}%
\altaffiliation{Present address: Fritz-Haber-Institut der MPG,
   Faradayweg 4-6, 14195 Berlin, Germany}%
\author{Roger E. Miller}%
\thanks{Deceased: Nov 6, 2005}%
\affiliation{Department of Chemistry, University of North Carolina at
   Chapel Hill, Chapel Hill, NC 27599}%

\date{\today}

\begin{abstract}
Rotationally resolved infrared spectra are reported for the X-HCN (X
= Cl, Br, I) binary complexes solvated in helium nanodroplets. These
results are directly compared with that obtained previously for the
corresponding X-HF complexes [J. M. Merritt, J. K\"upper, and R. E.
Miller, PCCP, 7, 67 (2005)]. For bromine and iodine atoms complexed
with HCN, two linear structures are observed and assigned to the
$^{2}\Sigma_{1/2}$ and $^{2}\Pi_{3/2}$ ground electronic states of
the nitrogen and hydrogen bound geometries, respectively.
Experiments for HCN + chlorine atoms give rise to only a single band
which is attributed to the nitrogen bound isomer. That the hydrogen
bound isomer is not stabilized is rationalized in terms of a
lowering of the isomerization barrier by spin-orbit coupling.
Theoretical calculations with and without spin-orbit coupling have
also been performed and are compared with our experimental results.
The possibility of stabilizing high-energy structures containing
multiple radicals is discussed, motivated by preliminary
spectroscopic evidence for the di-radical Br-HCCCN-Br complex.
Spectra for the corresponding molecular halogen HCN-X$_{2}$
complexes are also presented.
\end{abstract}


\maketitle

\section{Introduction}
Much of our knowledge of intermolecular forces has come from the
high-resolution spectroscopy of weakly bound closed shell
complexes \cite{110}, and there is considerable promise that
similar advances can be made for open shell complexes.
Experimentally, the study of open shell reactive molecular
complexes is more difficult than their closed shell counterparts
due to the fact that their tendency to react must also be
surpressed. The cooling provided by free jet expansion
\cite{13487,10758,15164} and matrix isolation
\cite{14697,14484,14009} is sometimes sufficient to stabilize
complexes of this type. Recently, nano-scale liquid helium
droplets have also emerged as a nearly ideal spectroscopic matrix
for the study of highly metastable species
\cite{10818,11603,14433,5971}, including the stabilization of pre-
and post-reactive cluster systems \cite{14522,14912,15251}. Such
weakly bound complexes sample the long range van der Waals forces
in the entrance and exit channel regions of the potential energy
surface, which have recently been shown to strongly influence the
reaction dynamics of Cl + HD \cite{14514,14281,15252} and the near
threshold photo-dissociation of formaldehyde \cite{14832,15239}.
Despite the fact that the magnitude of such orientational forces
are usually very small in comparison with the energy of the
transition state, the torque on the reactants at long range acts
to deflect trajectories towards or away from the transition state
\cite{14281}.

``Heavy+light-heavy'' systems such as X-HY, where X and Y are both
halogen atoms, have emerged as prototypical systems to study
chemical reaction dynamics at a fully quantum mechanical level
\cite{15253,9385,14015,15256,15254}.  The attraction of such
complexes is due to the fact that they contain only 2 heavy atoms,
thus simplifying the nuclear degrees of freedom, yet the
corresponding electronic degrees of freedom can be quite complex.
Indeed, because the unpaired electron of the free halogen atom
resides in a p orbital, three potential energy surfaces are needed
to fully characterize the interactions, which correspond to the
three relative orientations of the p orbital with respect to HX.
These surfaces become degenerate for large separations, and the
coupling between the surfaces cannot be neglected \cite{13847}.
Furthermore, spin-orbit coupling will also reshape the potential
energy surfaces \cite{13847,11590,15255}. In this work we extend
these studies by looking at a related X-HY complex, where Y is now
the pseudo-halogen CN.

The reactions of halogen atoms with HCN have been the focus of
extensive experimental \cite{12327,13257,5577} and theoretical
study \cite{13342,15240,15241}, aimed at elucidating the role of
the potential energy surface in determining the associated
reaction dynamics. The CN group is often regarded as a
pseudo-halogen, and thus there is interest in comparing the
reaction dynamics of this tetra-atomic system with the analogous
X-HY (X \& Y = halogen atoms) triatomic systems. Interestingly,
the long range interactions between XH + CN are expected to differ
significantly from X + HCN due to the threefold degeneracy
associated with the $^2P$ ground state of halogen atoms, compared
to the nondegenerate $^2\Sigma$ ground state of CN radicals,
further lowering the symmetry of this hydrogen exchange reaction.
Despite this, HCN is an ideal test case because the CN group is
usually found to act as a single unit, but there is still
considerable debate about whether or not the CN bond is a
spectator to the dynamics of many different reactions
\cite{15243,12458,5577}. The additional degrees of freedom that
the HCN molecule presents, open up many new channels of the
potential energy surface that can be explored, and the X + HCN
systems have provided benchmarks for the theoretical treatment of
such multi-dimensional dynamics.

The Cl-H-C-N potential energy surface has been studied
experimentally from three different starting positions. De Juan
\emph{et al.} \cite{15241} have investigated the H + ClCN
$\rightarrow$ HCl + CN reaction by colliding translationally hot H
atoms formed from the 248 nm photolysis of H$_{2}$S, with ClCN. This
experimental arrangement prepares the reactants with 21.6 kcal
mol$^{-1}$ of collision energy, which is $\sim$43 kcal mol$^{-1}$
above the energy of Cl + HCN (see Figure \ref{fig:X-HCN-G2}). By
analyzing the nascent CN rovibrational populations, as well as the
Doppler profiles of the vibrational transitions, it was proposed
that CN is a spectator under these conditions, and that most of the
available energy goes into HCl vibration and into translational
energy of the two fragments. Sims and Smith \cite{15243} and Frost
\emph{et al.} \cite{12458} have studied the HCl($v_{HCl}$) +
CN($v_{CN}$) $\rightarrow$ HCN + Cl reaction, and came to the
conclusion that CN is also a spectator in this reaction, due to the
fact that there is a negligible enhancement of the reaction rate
upon exciting the CN stretching vibration. A third set of
experiments were carried out by Metz \emph{et al.} \cite{5577,7706}
and Kreher \emph{et al.} \cite{9220,12327} focusing on the reaction
of translationally hot chlorine atoms with highly vibrationally
excited HCN:\[ Cl+HCN(v_{1},v_{2},v_{3})\rightarrow HCl+CN\] Metz
and coworkers showed that HCN, when prepared with either 4 quanta of
CH stretch (004) or 3 quanta of CN stretch plus two quanta of CH
stretch (302) react with comparable rates, and conclude that under
these conditions CN ``is clearly not a spectator''\footnote{The P(4)
transition of the (004) $\leftarrow$ (000) band (12623 cm$^{-1}$)
and the P(8) transition of the (302) $\leftarrow$ (000) band (12631
cm$^{-1}$) were used for the excitation}. Another interesting
observation from this work is that only 14\% of the available energy
was found as HCl vibration, which is surprisingly small since this
is the bond that is formed in the reaction. To justify this
observation an addition-elimination mechanism was proposed which
involved an intermediate HClCN product, as opposed to direct
collinear abstraction. Indeed, experiments in argon matrices have
now proven the existence of HFCN \cite{15155,15156}. If the
intermediate survived long enough in the gas-phase to permit energy
redistribution, this could account for the transfer of energy from
HCl($v$) to CN($v$). According to this mechanism the CN cannot be a
simple spectator. Kreher \emph{et al.} confirm these observations,
however point out that HCN(302) does give rise to higher vibrational
excitation of CN than does HCN(004), so some memory of the initial
state is retained.

The global potential energy surface for Cl + HCN was first studied
theoretically by de Juan \emph{et al.} at the MP4/3-21G{*} level
to aid in the interpretation of their experimental results
\cite{15241}. Their calculations did show the existence of a
covalently bound HClCN intermediate product, but the relative
energies of the reactants and products showed significant
discrepancies with the experimentally determined values. Harding
later revisited the Cl-HCN surface at the RHF+1+2+QC/cc-pVDZ level
and found the energies to be in much better agreement with
experiment \cite{15240}. Although the overall topology of the two
surfaces are similar, the existence of a direct HClCN
$\rightarrow$ HCl + CN pathway originally proposed by de Juan
\emph{et al.} was not found by Harding. Instead, Harding points
out that they most likely found the transition state for collinear
abstraction. Currently an exhaustive search of the six-dimensional
potential energy surface is not possible, however Harding states
that this HCl elimination pathway is unlikely based on the overall
topology of the potential. He finds that the transition state of
the Cl + HCN $\rightarrow$  HCl + CN reaction is a collinear
abstraction mechanism, and that the reaction path does not
directly involve the HClCN complex, in contrast to that proposed
by Metz and coworkers. Indeed the HClCN complex can be formed at
energies well below the abstraction reaction, however further
reaction to form HCl + CN would require surmounting barriers in
excess of that for direct abstraction. In order to explain the
experimental results, which clearly show vibrational excitation of
the CN, Harding proposes two possible mechanisms. First, the CN
stretching frequency is found in the calculations to be largely
unaffected along the reaction coordinate. In contrast, the reagent
CH stretch and product HCl stretches, which are initially above
the CN frequency, go through a minimum at the transition state and
are below the CN frequency. Because these normal modes cross the
CN stretching frequency, energy could flow between these two modes
along the course of the reaction. Secondly, Harding's calculations
predict the existence of long-range entrance channel complexes on
both sides of the abstraction transition state (Cl-HCN, HCN-Cl,
and CN-HCl) which could also influence the final product state
distributions.

Most recently Troya \emph{et al.} have performed a quasiclassical
trajectory study of the Cl + HCN $\rightarrow$ HCl + CN reaction
\cite{13342}. Their findings are in agreement with Harding, in that
the intermediate HClCN product only constitutes a minor channel
($<$5\%) at the translational energies relevant to the previous
experiments. The focus of the current study is to explore the
long-range entrance channel complexes of X + HCN.

\section{Experimental}
A detailed description of the experimental apparatus has been given
previously \cite{11280}, therefore only the most salient features
will be described here. Helium droplets are formed by expanding
ultra-high-purity helium gas through a 5 $\mu$m nozzle which is
cooled to 18-22 K.  The helium stagnation pressure was maintained at
50 bar, resulting in the formation of droplets with a mean size of
2500 - 6000 atoms \cite{10872}. Halogen radicals were produced by
pyrolysis \cite{13231,14522} and doped into the droplets using an
effusive source while the HCN is added to the droplets downstream
using a simple scattering cell. The documented growth of
non-equilibrium structures in helium droplets is suggestive of a
successive capture mechanism where dopants are pre-cooled before
complexation occurs \cite{10818,11603,12851}. The relative
timescales for species finding each other in a droplet, to form a
complex, and the rate of cooling are not precisely known. The
cooling rate however must at least be competitive with coagulation
in order to prevent annealing of previously formed metastable
structures in the pick-up process. In much the same way, the energy
gained due to the mutual interaction of two dopants must also be
quickly removed and complexation will occur at low interaction
energies comparable to the droplet temperature of 0.4 K. Therefore,
complexes are typically trapped in a local minimum reflecting the
approach geometry of the dopants; the low temperature does prevent
rearrangement to the global minimum \cite{10818,11603}.

The infrared light from an F-center Laser (Burleigh FCL-20) or a
PPLN-OPO (Linos Photonics OS-4000) interacts with the droplet beam
using a linear multipass cell, designed to increase the effective
interaction length. Excitation of the dopant inside the droplet
leads to evaporation of approximately 600 helium atoms from the
droplet \cite{5665}, which reduces the on-axis beam flux reaching
a liquid helium cooled bolometer. By mechanically chopping the
laser and using phase-sensitive detection, the resulting beam
depletion can be recorded as a function of the laser frequency to
obtain the absorption spectrum.  A set of Stark electrodes was
aligned orthogonal to the laser interaction so that Stark and
pendular spectroscopy could be performed. The DC electric field
was oriented parallel to the laser polarization yielding
$\Delta$M=0 selection rules. Details on the calibration of the
electric field are given elsewhere \cite{14143}. In order to
optimize conditions for the pickup of a single halogen atom, the
temperature of the pyrolysis source is adjusted, monitoring the
percent dissociation of the precursor by probing both the X-HCN
and X$_{2}$-HCN binary complexes.

\section{Electronic Structure calculations}
\begin{figure}
\includegraphics{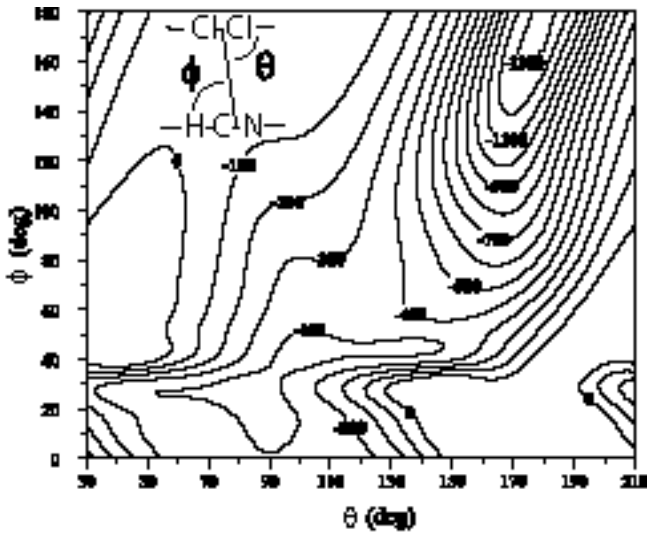}

\caption{\label{fig:HCN-Cl2-2DPES}A two-dimensional angular
potential for HCN-Cl$_{2}$ showing two possible local minimum
structures. At each point only the intermolecular distance was
optimized to find its minimum in energy. Counterpoise correction was
applied and the final surface generated using a bi-cubic spline
interpolation algorithm.  Contour lines are drawn in 100 cm$^{-1}$
intervals. In helium droplets both isomers are observed for I$_{2}$,
Br$_{2}$, and Cl$_{2}$ complexes with HCN, where as in the gas-phase
only the global minimum (linear nitrogen bound) isomer has been
identified \cite{13608}. Interestingly only the global minimum
HCN-F$_{2}$ isomer was observed in helium.}
\end{figure}

In order to characterize the precursor complexes we have calculated
a two-dimensional angular potential (at the RMP2/aug-cc-pVDZ level)
for HCN-Cl$_{2}$ corresponding to the two rotors being fixed within
the same plane. The two angles, $\theta$ and $\phi$, were stepped in
increments of 10\textdegree, and the intermolecular distance R was
relaxed to find its minimum in energy. The remaining geometric
parameters were held fixed at the values found for separate HCN and
Cl$_{2}$ optimizations, namely r$_{CH}$ = 1.0779 $\textrm{\AA}$,
r$_{CN}$= 1.1828 $\textrm{\AA}$, and r$_{ClCl}$ = 2.3275
$\textrm{\AA}$. The dissociation energy of the complex was corrected
for basis set superposition error \cite{4713}, and a spline
interpolation was used to generate the final surface shown in Figure
\ref{fig:HCN-Cl2-2DPES}.
 The surface predicts a deep minimum for a linear nitrogen bound geometry
($\theta$, $\phi$) = (180, 180), and a very flat minimum around
(90, 20) corresponding to a near T-shaped hydrogen bound complex.
Previous gas-phase studies \cite{13608,14561,14562} have confirmed
the linear geometry of the HCN-Cl$_{2}$, HCN-BrCl, and HCN-FCl
complexes however there has been no previous evidence (nor
prediction) of a hydrogen bound isomer. Fully relaxed geometry
optimizations and harmonic vibrational frequency calculations
confirm that the local minimum equilibrium structure is slightly
bent away from an exactly T-shaped geometry, however vibrational
averaging is likely to be important due to the weak anisotropy.
Exploratory calculations in which the restriction of planarity was
relaxed did not result in any new minima.

\begin{figure*}
\includegraphics[width=6in]{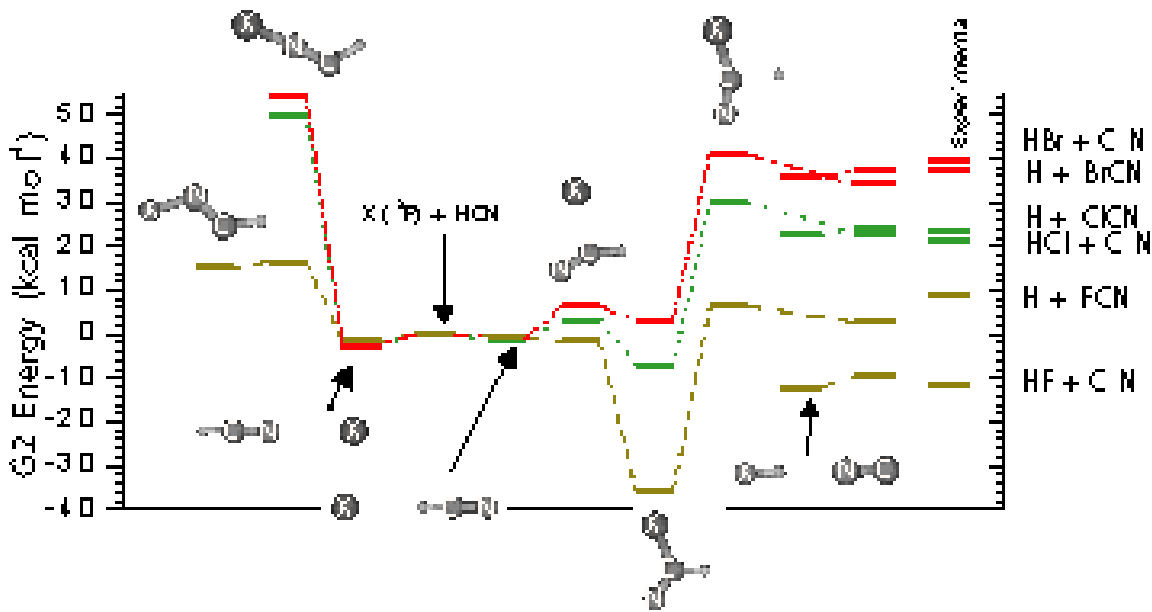}

\caption{\label{fig:X-HCN-G2} Relevant stationary points of the X +
HCN reactions calculated using the composite G2 method. Two reacted
complexes in addition to two weakly bound entrance channel species
are predicted. The HCNX products of chlorine and bromine atoms
binding on the nitrogen end of HCN could not be stabilized in our
calculations, however HCNCl was found in reference \cite{15240}. The
experimental heats of formation for the X + HCN $\rightarrow$ HX +
CN and X + HCN $\rightarrow$ H + XCN reactions are taken from
reference \cite{12826}. The photon energy used to study the
pre-reactive X-HCN complexes is greater than the barriers to form
HXCN, suggesting that it may be possible to photo-initiate this
reaction. }
\end{figure*}

As pointed out by Harding \cite{15240}, the reactivity toward HCN
by halogen atoms is such that two reaction products have been
postulated, with the halogen covalently binding to either the
carbon or nitrogen atom of HCN. Due to limitations of the basis
set used and computational expense, Harding admits that the
calculations performed to date are more qualitative in nature. In
this section we extend these calculations to higher-levels of
theory and calculate each of the F, Cl, and Br reactions so that
they may be directly compared, but we only focus on the lowest
energy channels including HXCN formation. To explore the relative
energetics of these complexes, we have employed the G2 method
\cite{14497}, and the resulting stationary points, including
zero-point energy, on the potential energy surface are shown
graphically in Figure \ref{fig:X-HCN-G2}.
 For comparison, a few stationary points have also been calculated
at the UMP2/aug-cc-pVTZ and UCCSD(T)/6-311++G(d,p) levels in
addition to the G2 results, and are summarized in Table
\ref{table:X-HCN-reactions}.
\begin{table*}\begin{tabular}{cccccc}
\hline & G2& UCCSD(T) & UMP2 & Experimental & Harding
\cite{15240}\tabularnewline & & 6-311++G(d,p)& aug-cc-pVTZ& ($\Delta
H$)& RHF+1+2+QC/cc-pVDZ\tabularnewline \hline \hline Cl (F) + HCN&
0.0& 0.0& 0.0& 0.0& 0.0\tabularnewline
 T.S.&
3.30 (-1.70)& 6.97 (3.05)& -$^{a}$& -& 8.2\tabularnewline HClCN
(HFCN)& -7.07 (-35.83)& -3.41 (-28.44)& 4.91 (-27.18)& -&
-2.2\tabularnewline HCNCl (HCNF)& -$^{a}$ (15.32)& -$^{a}$ (22.60)&
-$^{a}$ (26.86)& -& 27.9\tabularnewline HCl (HF) + CN& 24.21
(-9.45)& 25.37 (-6.54)& 42.90 (6.18)& 21.41 (-11.58)&
25.9\tabularnewline ClCN (FCN) + H& 23.11 (2.71)& 26.80 (9.17)&
19.03 (-3.03)& 23.45 (9.26)& 28.8\tabularnewline \hline
\multicolumn{5}{l}{$^{a}$Could not be stabilized in the geometry
optimization.}& \tabularnewline
\end{tabular}

\caption{\label{table:X-HCN-reactions}A summary of the calculated
energetics of the Cl (F) + HCN reactions including zero-point
energy. T.S. corresponds to the transition state of the reaction
with the carbon atom of HCN to form HXCN. The experimentally
determined heats of reaction ($\Delta H$) are also shown for
reference \cite{12826}. One can see that while the G2 and UCCSD(T)
methods predict the enthalpies quite well, the UMP2 method shows
much larger deviations. A graphical representation of these
energies is shown in Figure \ref{fig:X-HCN-G2} for the G2
calculations. Previous calculations by Harding for Cl + HCN are
also given for reference.}
\end{table*}
The calculated enthalpies of reaction are found to be in
qualitative agreement with experiment at the G2 and
UCCSD(T)/6-311++G(d,p) levels of theory, however the agreement is
much worse at the UMP2 level.

From Figure \ref{fig:X-HCN-G2} it is clear that the X + HCN
$\rightarrow$ HCNX reactions are all quite endothermic, and
therefore we expect that this reaction channel will not play at
 role in the dynamics at 0.37 K. More relevant to
the experimental conditions is the reactivity of the halogen atoms
with the carbon atom of HCN. For the F + HCN reaction, a very stable
HFCN intermediate product is found in the calculations. At the G2
level no barrier is predicted for this reaction, however at the
UCCSD(T)/6-311++G(d,p) level the reaction barrier is 3.05 kcal
mol$^{-1}$. Four of the six vibrational modes of this product were
first observed by Andrews \emph{et al.} \cite{15155} in an argon
matrix by codepositing F$_{2}$ and HCN, and then subjecting the
matrix to broadband UV photolysis. Although rotational resolution
was not achieved, the observation of the CN stretching vibration at
1672 cm$^{-1}$ indicates that the CN bond has double-bond and not
triple-bond character. Of interest to the current study, the CH
stretching vibration was observed at 3016 cm$^{-1}$. Later Misochko
\emph{et al.} confirmed Andrews observations, and assigned the two
remaining vibrational modes which were not observed previously
\cite{15156}. Experimental evidence for a pre-reactive X+HCN complex
or HClCN however is still lacking.

Harding's calculations on Cl+HCN are given in Table
\ref{table:X-HCN-reactions}, and in general, agree well with our
current calculations. G2 (UCCSD(T)) calculations predict that the
reaction of a chlorine atom with the carbon of HCN is exothermic by
-7.07 (-3.41) kcal mol$^{-1}$ , and that the barrier height is 3.30
(6.97) kcal mol$^{-1}$ . For bromine atoms, the reaction to form
HBrCN is predicted to be endothermic at the G2 level of theory.
Since the CH stretching vibration of the entrance channel complexes
has a higher frequency than the barrier to form HXCN, it may be
possible to photo-initiate the reaction starting from the entrance
channel.

\begin{figure*}
   \centering%
   \includegraphics[width=6in]{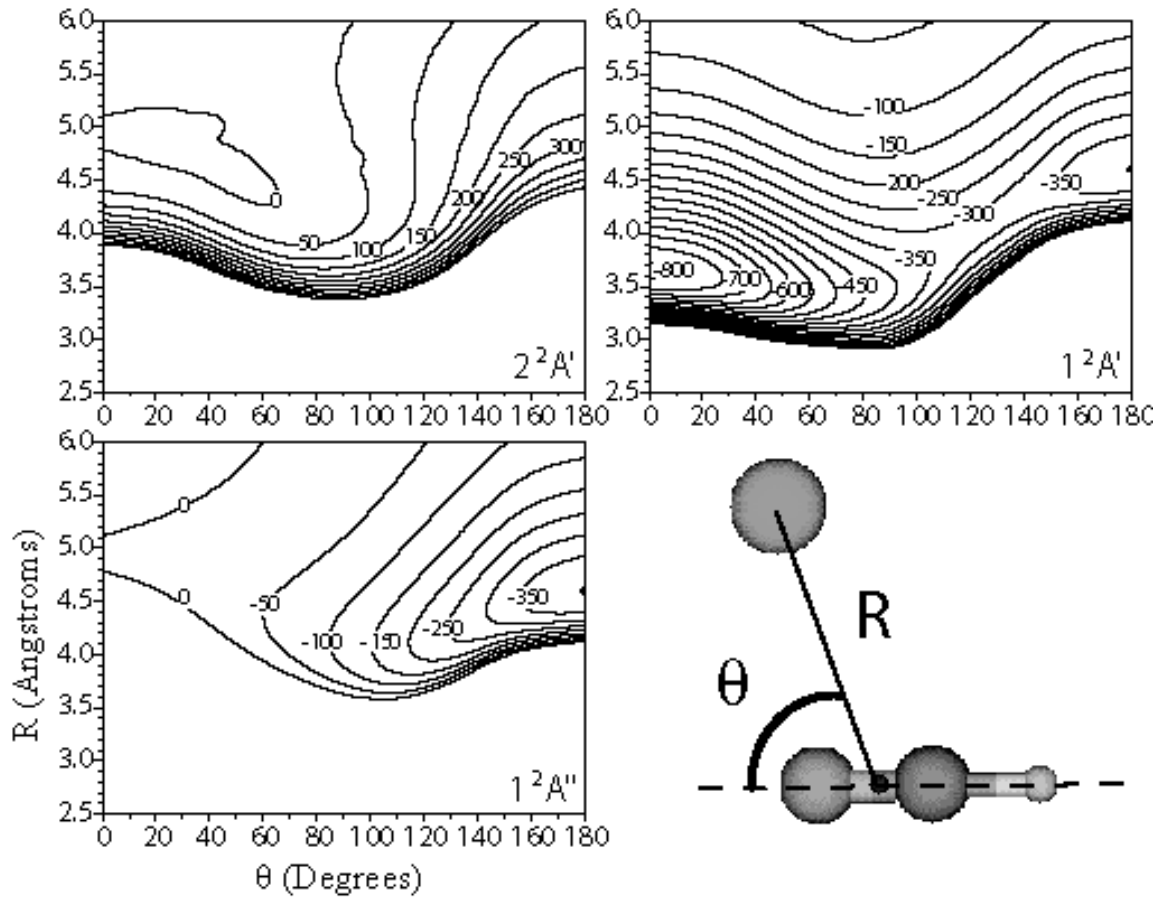}%
   \caption{The three non-relativistic adiabatic potential energy
      surfaces for HCN + Br calculated at the
      RCCSD(T)/aug-cc-pVDZ+\{332\} level. Contour lines are drawn in 50 cm$^{-1}$
intervals. The surfaces predict both
      linear structures to be minima.}%
   \label{fig:HCN-Br-2DPES}%
\end{figure*}

While the above calculations give us an overview of the reaction
energetics for the halogen atom - HCN reactions, we are
particularly interested in the topology of the surface around the
entrance channel. For instance, it appears that HCN+X might be
most reactive in a T-shaped arrangement, so the long-range
orientational effects of the potential may be very important in
determining the reaction dynamics. To begin to answer these
questions we have calculated three-dimensional (R, $\theta$,
r$_{CH}$) non-relativistic adiabatic potential energy surfaces for
Br + HCN and Cl + HCN at the RCCSD(T)/aug-cc-pVDZ+\{332\} level of
theory. The transformation of these surfaces to include spin-orbit
coupling, and the calculation of the resulting bound states is
being performed in collaboration with the group of van der Avoird,
and details on these results will be presented elsewhere
\cite{15248}. In order to gain insight into our experimental
results, we will focus our attention on the qualitative features
of a 2D slice (r$_{CH}$ = 1.0655 $\textrm{\AA}$) from our 3D
non-relativistic adiabatic potentials, which are shown in Figure
\ref{fig:HCN-Br-2DPES} for Br + HCN. The incorporation of the
dimension corresponding to the CH bond length of HCN (r$_{CH}$)
will lend insight into the change of the 2D potentials upon
vibrational excitation  of HCN ($\nu_1$).

Calculations were performed using MOLPRO \cite{13953}, employing
Jacobi coordinates, $R$ and $\theta$, and restricting the
electronic wavefunction to C$_{s}$ symmetry, which gives rise to
the 1A', 2A', and 1A'' surfaces, corresponding to the three
relative orientations of the unpaired orbital of the halogen atom
with respect to HCN. A set of uncontracted mid-bond functions with
exponents sp: 0.9, 0.3, 0.1, and d: 0.6, 0.2 (denoted as \{332\})
have been added to the aug-cc-pVDZ basis set and were placed at
the midpoint between the halogen atom and the HCN nuclear center
of mass \cite{14016,15193}. For the plots shown in Figure
\ref{fig:HCN-Br-2DPES}, R and $\theta$ have been incremented by
0.1 $\textrm{\AA}$ and 10\textdegree \ respectively. The r$_{CH}$
and r$_{CN}$ bond lengths of HCN were fixed at 1.0655 and 1.1532
$\textrm{\AA}$ respectively. Basis set superposition error has
been accounted for by performing counterpoise correction
\cite{4713}, and a spline interpolation used to smooth the
surface. As noted by Fishchuk \emph{et al.} for Cl-HF
\cite{15189}, there is some choice in how counterpoise correction
is applied due to the two A' states of the free halogen atom. Our
procedure is the same as that reported for Cl-HF, in that the
lowest RCCSD(T) energy of the free halogen atom was subtracted
from both A' dimer energies, which preserves the double degeneracy
of the $\Pi$ state for both linear geometries. The nitrogen bound
(HCN-X) geometry corresponds to $\theta$ = 0\textdegree.

The surfaces clearly show that both linear geometries have stable
minima, with well depths of -810 and -401 cm$^{-1}$ for the HCN-Br
and Br-HCN isomers respectively. The isomerization barrier
connecting Br-HCN to HCN-Br is found to be only 90 cm$^{-1}$
(neglecting zero-point effects). At $\theta$ = 0, the 1A'' and 2A'
surfaces correlate with the doubly degenerate $^{2}\Pi$ state, where
as for $\theta$ = 180, it is the 1A' and 1A'' surfaces. The ground
electronic state of the HCN-Br isomer is $^{2}\Sigma$ and for Br-HCN
it is $^{2}\Pi$ in agreement with that found by Harding
\cite{15240}. These symmetries illustrate that the unpaired orbital
of the halogen atom is along the axis of the molecule for
$^{2}\Sigma$, while it is perpendicular to the molecular axis for
$^{2}\Pi$. These different electronic symmetries are determined by
electrostatic interactions of this dipole-quadrupole system. In the
linear nitrogen bound complex the negative end of the HCN dipole
will want to approach the positive p-hole created by the unpaired
orbital, thus orienting the unpaired orbital on the molecular axis.
When the positive end of the HCN dipole approaches the Br, the
positive p-hole will be repelled, and orient itself perpendicular to
the molecular axis, resulting in a $\Pi$ ground state.

Fully relaxed geometry optimizations, at the RMP2 level, were also
undertaken for each of these stable structures to generate dipole
moments and scaled harmonic frequencies to compare with our
experiment, which are summarized in Table \ref{table:HCN-X-MP2}.
Note that vibrational frequencies could not be calculated for the
basis sets which include mid-bond functions, so we have instead
used the aug-cc-pVTZ basis set. Calculations for iodine containing
complexes employed the aug-cc-pVTZ-PP \cite{14349}
pseudo-potential for the iodine atom. One can see that while the
binding energy for Br-HCN is in very good agreement with the 2D
surfaces (399.5 cm$^{-1}$ at the RMP2/aug-cc-pVTZ level compared
with the 2D RCCSD(T)/aug-cc-pVDZ+\{332\} result of 401 cm$^{-1}$),
the differences found for HCN-Br are somewhat greater (910 vs. 810
cm$^{-1}$).

\begin{table*}
\begin{tabular}{ccccc} \hline RMP2& HCN-F& HCN-Cl& HCN-Br& HCN-I
(ECP)\tabularnewline aug-cc-pVTZ& (F-HCN)& (Cl-HCN)& (Br-HCN)&
(I-HCN (ECP))\tabularnewline \hline \hline D$_{e}$ (cm$^{-1}$)&
405.2& 834.3& 909.8& 935.1\tabularnewline & (189.0)& (359.9)&
(399.5)& (429.4)\tabularnewline $\nu_{harmonic}$ (cm$^{-1}$)&
-$^{a}$& 3464.21& 3463.11& -$^{a}$\tabularnewline & -$^{a}$&
(3443.85)& (3433.52)& (3430.88)\tabularnewline $\nu_{scaled}$
(cm$^{-1}$)& -& 3309.11& 3308.06& -\tabularnewline & -& (3291.66)&
(3281.78)& (3279.26)\tabularnewline R ($\textrm{\AA}$)& 3.240&
3.416& 3.542& 3.726\tabularnewline & (4.015)& (4.342)& (4.475)&
(4.670)\tabularnewline B (cm$^{-1}$)& 0.1310& 0.0885& 0.0636&
0.0525\tabularnewline & (0.0881)& (0.0561)& (0.0405)&
(0.0339)\tabularnewline $\mu$ (D)& 3.24& 3.66& 3.77&
3.91\tabularnewline & (3.18)& (3.40)& (3.47)&
(3.58)\tabularnewline \hline \multicolumn{5}{l}{$^{a}$Tight
geometry convergence was achieved but imaginary frequencies were
calculated}\tabularnewline
\end{tabular}

\caption{\label{table:HCN-X-MP2}Computed properties of the HCN-X
(X-HCN) isomers at the RMP2/aug-cc-pVTZ level. The calculations
for iodine incorporate a small core relativistic pseudo-potential
(aug-cc-pVTZ-PP) in which 28 core electrons were replaced with the
ECP \cite{14349}. Binding energies have been corrected for BSSE
 \cite{4713}. Harmonic frequencies have been scaled by a factor of 0.9552,
which is obtained by comparing a HCN calculation at this level to
the corresponding experimental helium droplet band origin. R, is
the X-HCN center of mass distance.}
\end{table*}

\subsection{Relativistic - adiabatic potential energy surfaces\label{sub:X-HCN-relativistic}}

The effects of spin-orbit coupling on reshaping the potential
energy surface is well documented \cite{14016,12216} and in order
to make detailed comparisons with our experiment we must
incorporate it. To include spin-orbit coupling for the linear
geometries (1D slices) of our 2D \emph{ab initio} potentials we
take note of the operator form of the spin-orbit
Hamiltonian:\begin{eqnarray*}
H^{so} & = & A\textbf{L}\cdot \textbf{S}\\
 & = & A(L_{z}S_{z}+\frac{L^{+}S^{-}+L^{-}S^{+}}{2})\end{eqnarray*}
The spin-orbit coupling constant, A, is -269.3, -587.3, -2457, and
-5068 cm$^{-1}$ for F, Cl, Br, and I, respectively \cite{1228}.
Note that the negative sign for A illustrates that the
$^{2}P_{3/2}$ state lies below the $^{2}P_{1/2}$ state in the free
atoms. We make the approximation that the spin-orbit coupling is
independent of the geometry of the complex, and remains that of
the isolated atom. That this approximation is valid for the
long-range part of our potentials is justified because the
electron-nuclei interaction scales as $\left\langle
\frac{1}{r^{3}}\right\rangle $\cite{1228}, and therefore the inner
lobes of the orbitals are more heavily weighted, which are
expected to be only slightly perturbed by the relatively weak van
der Waals interaction. Recent calculations on O($^{3}P$) + HCl by
Rode \emph{et al.} \cite{15195} have examined the validity of this
approximation by explicit calculation of the spin-orbit matrix
elements using the Breit-Pauli operator \cite{15210}. Their
results indicate that for internuclear distances similar to the
van der Waals separation the spin-orbit coupling is not
substantially affected.

\begin{figure}
\includegraphics[width=\linewidth]{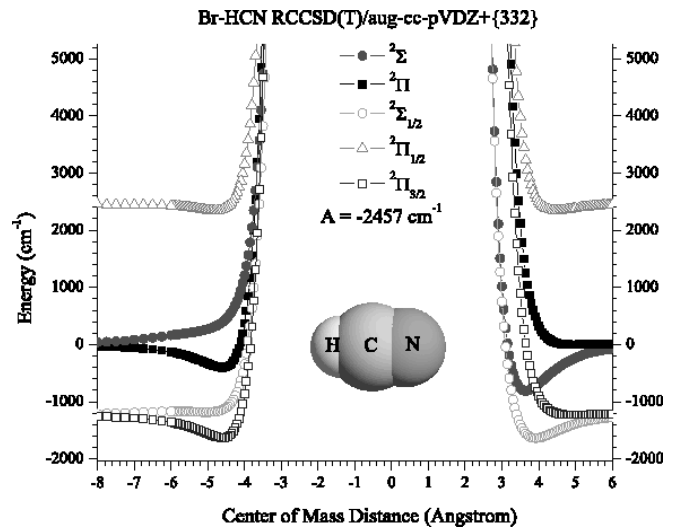}

\caption{\label{fig:HCN-Br-rel}Relativistic and non-relativistic
adiabatic potential energy curves for the HCN-Br complex derived
from the \emph{ab initio} results. See text for details.}
\end{figure}

The diagonal term of the spin-orbit Hamiltonian ($L_{z}S_{z}$)
gives the $^{2}\Pi_{1/2}$ and $^{2}\Pi_{3/2}$ potentials as
$^{2}\Pi\mp\frac{A}{2}$, while the $^{2}\Sigma$ potential is
simply $^{2}\Sigma_{1/2}$. This is illustrated by plugging in the
eigenvalues of the L$_{z}$ and S$_{z}$ operators, namely
$\Lambda=1$ and $\Sigma=+1/2$ for the $^{2}\Pi_{3/2}$ state and
$\Lambda=0$ and $\Sigma=\left|1/2\right|$ for the
$^{2}\Sigma_{1/2}$ state. The off-diagonal spin-orbit term
($L^{\pm}S^{\mp}$) couples the $^{2}\Pi_{1/2}$ state with the
$^{2}\Sigma_{1/2}$ state and has the following form \cite{1228}:

\begin{eqnarray}
\left\langle
^{2}\Pi_{1/2}|\frac{A}{2}L^{+}S^{-}|^{2}\Sigma_{1/2}\right\rangle
=2^{1/2}(A/2) \nonumber \end{eqnarray} To derive the relativistic
adiabatic potentials we set up a R dependent 2 $\times$ 2 matrix
for the $\Omega=1/2$ states, set the diagonal elements to the
energies from the \emph{ab initio} calculation, and the
off-diagonal elements to $2^{1/2}(\frac{A}{2})$, and diagonalize
the matrix numerically to get its eigenvalues and eigenvectors.
For this simple two level system, the potentials (eigenvalues) are
given by:\begin{eqnarray}
^{2}\Sigma_{1/2} & = & a^2(^{2}\Pi-\frac{A}{2})+(1-a^{2})(^{2}\Sigma)\nonumber \\
^{2}\Pi_{1/2} & = & a^2(^{2}\Sigma)+(1-a^{2})(^{2}\Pi-\frac{A}{2})
\nonumber \end{eqnarray} where $a$ is an R dependent mixing
coefficient. The remaining $^{2}\Pi_{3/2}$ state is simply
$^{2}\Pi+\frac{A}{2}$. The diagonalized relativistic and
non-relativistic adiabatic potentials are shown in Figure
\ref{fig:HCN-Br-rel}.
At long range, where the $\Sigma$ and $\Pi$ states are nearly
degenerate, the $^{2}\Sigma_{1/2}$ state is pushed lower in energy
compared to $^{2}\Sigma$ while the $^{2}\Pi_{1/2}$ is pushed to
higher energies compared to $^{2}\Pi-A/2$ due to the coupling. In
the limit of R $\rightarrow\infty$ , the $^{2}\Pi_{3/2}$ and
$^{2}\Sigma_{1/2}$ states converge and the separation between
$^{2}\Pi_{1/2}$ and $^{2}\Pi_{3/2}$ becomes (3/2)A, or that observed
in the free atom. The bound states of these potentials were
calculated numerically using the program Level \cite{15165} and the
binding energies, rotational constants, and the van der Waals
stretching frequencies are summarized in Table
\ref{table:HCN-Br-bound-state-properties}. To estimate the effects
of CH stretching of HCN on the potentials, computations were also
performed in which the HCN geometry was constrained to have a CH
bond length of r$_{CH}$ = 1.09 $\textrm{\AA}$, which is close to the
vibrationally averaged bond length for the $v=1$ state. In general,
the excitation of HCN results in a slight increase in the binding
energy of the complex which we attribute to an increased dipole -
quadrupole interaction.
\begin{table*}
\begin{tabular}{cccc} Constant (cm$^{-1}$)& & Br-HCN&
\tabularnewline HCN $v_{1}=0$ ($v_{1}=1$)& $^{2}\Pi$&
$^{2}\Pi_{1/2}$& $^{2}\Sigma_{1/2}$\tabularnewline \hline \hline
D$_{0}$& 376.43 (390.04)& $^{a}$& $^{a}$\tabularnewline B$_{0}$&
0.03893 (0.03893)& $^{a}$& $^{a}$\tabularnewline $\nu_{vdws}$&
47.37 (48.36)& $^{a}$& $^{a}$\tabularnewline & & & \tabularnewline
& & HCN-Br& \tabularnewline & $^{2}\Sigma$& $^{2}\Sigma_{1/2}$&
$^{2}\Pi_{1/2}$\tabularnewline D$_{0}$& 773.23 (775.89)& 383.33
(384.94)& 111.24\tabularnewline B$_{0}$& 0.06233 (0.06235)&
0.05379 (0.05382)& 0.04051\tabularnewline $\nu_{vdws}$& 69.71
(69.84)& 47.88 (47.97)& 26.91\tabularnewline \hline
\multicolumn{4}{l}{$^{a}$No bound states found}\tabularnewline
\end{tabular}

\caption{\label{table:HCN-Br-bound-state-properties}A summary of the
computed properties for the HCN-Br and Br-HCN isomers from bound
state calculations on one-dimensional relativistic adiabatic
potentials. Note that the computed bound states for the
$^{2}\Pi_{3/2}$ state are the same as those for the non-relativistic
$^{2}\Pi$ state.}
\end{table*}

 As noted above, spin-orbit coupling can strongly influence the potentials,
and indeed the dissociation energy of the HCN-Br
($^{2}\Sigma_{1/2}$) complex is much smaller than in the
non-relativistic calculation (383 vs 773 cm$^{-1}$). The
spin-orbit interaction is even strong enough to create a bound
state in the upper $^{2}\Pi_{1/2}$ well, despite the fact that the
$^{2}\Pi$ state is purely repulsive. But since this state
correlates with the excited spin-orbit component of the atom, we
would not expect this isomer to be produced in helium. For the
hydrogen bound isomer, the ground state is predicted to be
$^{2}\Pi_{3/2}$. Since this potential is not distorted from the
non-relativistic $^{2}\Pi$, we expect that the nonrelativistic
\emph{ab initio} calculations for this isomer should be sufficient
to compare with experiment. This may not be the case for the HCN-X
complexes, and thus the nonrelativistic \emph{ab initio} results
should be interpreted with caution. After inclusion of the
spin-orbit coupling, we find that both isomers for bromine are now
nearly isoenergetic. The smaller spin-orbit coupling constant for
chlorine atoms allows the HCN-Cl complex to remain the global
minimum. We look forward to the results of including spin-orbit
coupling into our 2D and 3D surfaces as it will be particularly
interesting to investigate how the isomerization barrier is
effected.
\begin{turnpage}
\begin{table*}
\begin{tabular}{ccccccc}
\hline Constant (cm$^{-1}$)& \multicolumn{2}{c}{HCN-$^{35}$Cl}&
\multicolumn{2}{c}{HCN-$^{79}$Br}&
\multicolumn{2}{c}{HCN-I$^{a}$}\tabularnewline v=0 (v=1)&
$^{2}\Sigma$& $^{2}\Sigma_{1/2}$& $^{2}\Sigma$&
$^{2}\Sigma_{1/2}$& $^{2}\Sigma$&
$^{2}\Sigma_{1/2}$\tabularnewline \hline \hline D$_{0}$& 687.65
(689.98)& 476.78 (478.86)& 773.23 (775.89)& 383.33 (384.94)&
781.08& 375.61 (377.24)\tabularnewline B$_{0}$& 0.08784 (0.08786)&
0.08478 (0.08482)& 0.06233 (0.06235)& 0.05379 (0.05382)& 0.05038&
0.04245 (0.04247)\tabularnewline $\nu_{vdws}$& 75.85 (75.98)&
64.23 (63.53)& 69.71 (69.84)& 47.88 (47.97)& 62.37& 43.44
(43.52)\tabularnewline $A$& -& -587.3& -& -2457& -&
-5068\tabularnewline \hline \multicolumn{7}{l}{$^{a}$This
calculation was performed with the aug-cc-pVDZPP+\{332\} basis set
\cite{14349}.}\tabularnewline
\end{tabular}
\caption{\label{table:HCN-X-1D-properties}A summary of the molecular
   parameters derived from the bound states of one-dimensional
   potential energy surfaces (see text). $A$ is the atomic spin-orbit
   coupling constant used in the calculations and $\nu_{vdws}$ is the
   van der Waals stretching frequency, defined as the energy separation
   between the two lowest calculated bound states. For the ``$v=1$''
   calculations, the CH bond length was fixed at 1.09 $\textrm{\AA}$
   (compared to 1.0655 $\textrm{\AA}$ for $v=0$) in order to estimate
   the effects of vibrational excitation of the HCN. Note that the
   corresponding calculations for the X-HCN isomers are not presented
   since the ground electronic state of these complexes is
   $^{2}\Pi_{3/2}$, which does not experience a first order coupling
   with the $^{2}\Sigma$ state, and so the molecular parameters are the
   same as those found in a standard \emph{ab initio} calculation
   presented in Table \ref{table:HCN-X-MP2}.}
\end{table*}
\end{turnpage}
Table \ref{table:HCN-X-1D-properties} collects the molecular
parameters derived from the bound states for the HCN-Cl, HCN-Br, and
HCN-I complexes.

\section{The HCN-X$_{2}$ complexes}

Given that previous experimental \cite{14562,14561} studies have
shown that halogen molecules bind to the nitrogen end of HCN, the
search for the HCN-X$_{2}$ complexes was straightforward given the
weak perturbation on the CH stretching frequency.
\begin{figure*}
\includegraphics[width=6in]{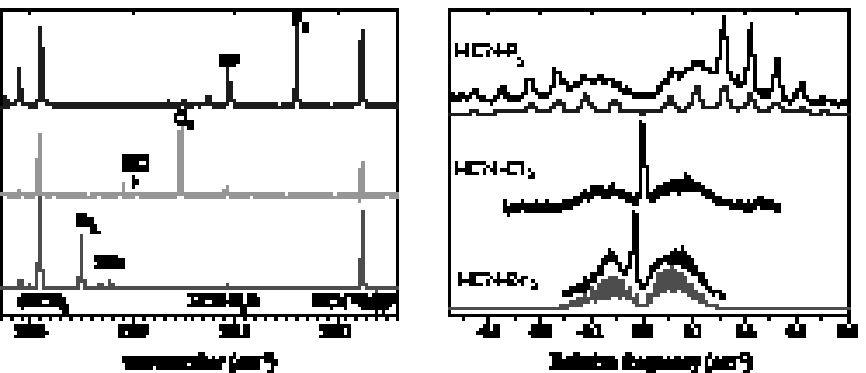}

\caption{\label{fig:HCN-X2-pendular-survey and FF}Pendular survey
scans showing the formation of HCN-X$_{2}$ molecular complexes (left
panel) and their corresponding field-free spectra (right panel).
Significant perturbations are observed in the HCN-Br$_{2}$ and
HCN-Cl$_{2}$ spectra (leading to the otherwise forbidden Q branch)
which we ascribe to a helium droplet interaction.}
\end{figure*}
Pendular survey scans for Br$_{2}$, Cl$_{2}$, and F$_{2}$ + HCN are
shown in Figure \ref{fig:HCN-X2-pendular-survey and FF}, and the new
peaks are labeled according to their gas dependence. The smooth
variation in frequency shifts is telling of the magnitude of the
interactions. Figure \ref{fig:HCN-X2-pendular-survey and FF} also
shows the resulting zero-field spectra for the HCN-Br$_{2}$,
HCN-Cl$_{2}$, and HCN-F$_{2}$ binary complexes. Molecular iodine
complexes with HCN were also observed but are not shown in the
figure. Due to the fact that our calculations predict that each
complex is linear, in accordance with the previous microwave results
\cite{14562,14561}, the Q branches in the bromine and chlorine
spectra are clearly unexpected. At this time we leave their
explanation to a future study. We only postulate here that an
impurity complex could be overlapping the band, or that the helium
interaction with these species is quite strong, due to the increased
polarizability compared to F$_{2}$, and thus more strongly
influences the spectra \cite{11279,13829}. A similar effect in HCN
trimer, another linear complex with a Q branch, has been treated
theoretically, and the Q branch was found to result from the thermal
excitation of the droplet in the first few solvation shells, which
rotates more rigidly about the \emph{a}-axis \cite{13829}. Possibly
these molecular halogen complexes exhibit similar behavior.

\begin{figure}
\includegraphics[width=0.7\linewidth]{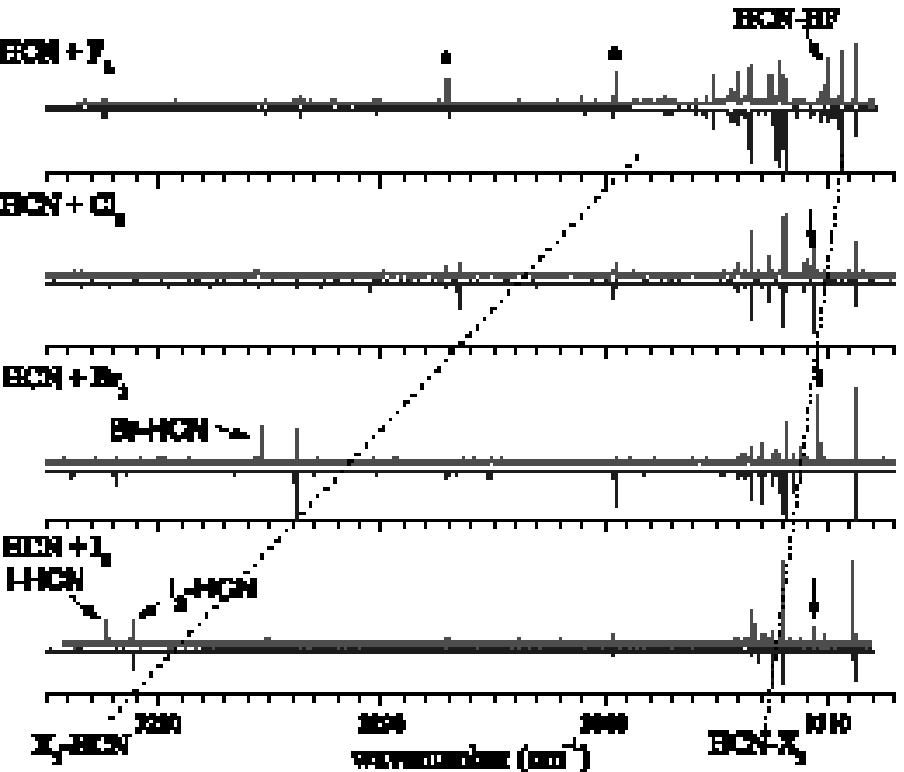}

\caption{\label{fig:X-X2-HCN-pendular-survey}Pendular survey scans
revealing the pyrolysis source temperature dependence (upward peaks
are with a hot source, downward peaks are with a cold source) on the
corresponding HCN + halogen experiment. The scans show peaks which
we identify as molecular and atomic halogens complexed with HCN. The
dotted lines are drawn to guide the eye, exemplifying the linear
scaling of the molecular halogen complex frequency shifts. The peaks
labeled with an asterisk are known impurities. Vertical arrows point
to the peaks which correspond to free CH stretch HCN-X complexes.}
\end{figure}
Longer pendular scans were also carried out to search the hydrogen
bonded frequency region, given the prediction of a second, nearly
T-shaped, minimum. Indeed, as shown in Figure
\ref{fig:X-X2-HCN-pendular-survey} for the case of bromine, a
strong peak is observed at 3286.14 cm$^{-1}$ which we assign to
such a complex. Our assignment is based on pick-up cell pressure
dependence measurements (optimized 1:1 with HCN-Br$_{2}$), scaled
harmonic frequency calculations, rotational band contour analysis,
and pyrolysis source temperature dependence. Hydrogen bonded
complexes between molecular iodine and chlorine with HCN were also
observed whereas the corresponding isomer of fluorine was not. The
survey scans for all cases are shown in Figure
\ref{fig:X-X2-HCN-pendular-survey} as downward going peaks. The
fact that a hydrogen bound F$_{2}$-HCN isomer is not formed might
be attributed to a smaller well depth, due to the smaller
polarizability of fluorine compared with chlorine, bromine, and
iodine, and larger zero-point energy effects allowing it to
convert back to the global minimum linear nitrogen bound geometry,
which is observed. The infrared spectra for these complexes taken
under field-free conditions will be presented elsewhere. The
dotted lines in Figure \ref{fig:X-X2-HCN-pendular-survey}
illustrate the almost perfect linear scaling of the band origins
for both the free stretch and the hydrogen bonded complexes.

\section{The B\lowercase{r}-HCN and I-HCN complexes}

\subsection{Br-HCN}

Given the identification of the precursor complexes, we are now in
a position to heat the pyrolysis source to produce the
corresponding atoms. As in our previous study of the X-HF systems,
at sufficiently high pyrolysis source temperatures the signal
levels associated with the molecular halogen complexes decrease,
due to dissociation into atoms, and a complementary set of peaks
grow in. The pendular survey scans performed with a hot pyrolysis
source are shown in Figure \ref{fig:X-X2-HCN-pendular-survey}, as
upward going peaks. Each scan was taken at the appropriate
pyrolysis source temperature for the dissociation of that
particular precursor. Concentrating on the bromine spectra, two
new peaks are observed with the hot source, namely at 3309.55 and
3284.61 cm$^{-1}$. Based on the frequency shifts of these two
bands it is already apparent that both linear isomers of HCN - Br
may be formed, with the hydrogen bonded isomer giving a much
larger redshift on the CH stretching frequency than the
corresponding nitrogen bonded isomer. We choose to first focus on
the peak we preliminarily assign to a hydrogen bonded complex
since a direct comparison can be made with the corresponding X-HF
complexes studied previously \cite{14522}.

\begin{figure*}
\includegraphics{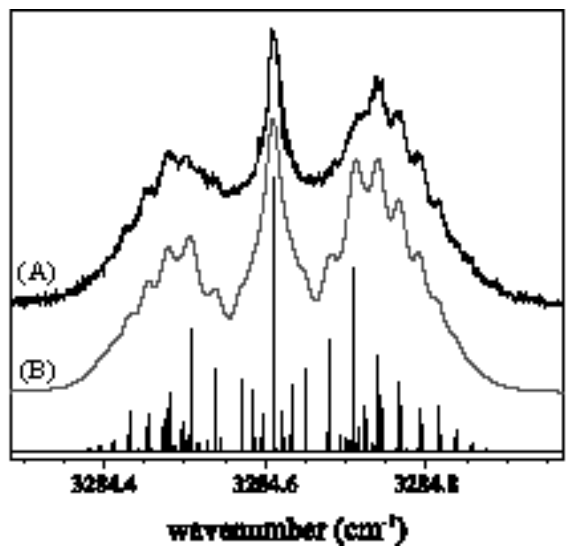}

\caption{\label{fig:Br-HCNspec-fit} (A) The rotationally resolved
spectrum of the hydrogen bound Br-HCN isomer recorded under electric
field-free conditions. The band shape is consistent with a
$^{2}\Pi_{3/2}$ ground electronic state which is in agreement with
theoretical predictions. The simulation (B) includes a nuclear
magnetic hyperfine interaction (I=3/2 for bromine) due to the large
magnetic moment of the bromine nucleus. A Lorentzian linewidth of
0.024 cm$^{-1}$ was used for the simulation. }
\end{figure*}

A field-free spectrum of the pendular peak centered at 3284.61
cm$^{-1}$ is shown in Figure \ref{fig:Br-HCNspec-fit}(A).
Again this band is only observed under conditions appropriate for
bromine atom pickup. In agreement with that found for the Br-HF
complex, P, Q, and R branches are observed, consistent with a
linear complex in a $\Pi$ ground electronic state. For Br-HF a
distinct 5B gap was observed between the Q branch transitions and
those arising from the first P and R branch transitions which
confirmed the ground state is $^{2}\Pi_{3/2}$, consistent with the
fact that the ground state of free bromine atoms is $^{2}P_{3/2}$.
The smaller rotational constant of Br-HCN however, makes these
gaps in the spectra more obscure, but due to the predicted weak
coupling of the spin-orbit interaction upon complexation, we can
say with confidence that this is also $^{2}\Pi_{3/2}$ band in
analogy with Br-HF. A fit to the spectrum adopting the same
Hamiltonian as used for Br-HF is shown in Figure
\ref{fig:Br-HCNspec-fit}(B). A nuclear magnetic hyperfine
perturbation of the form \cite{14306}:

\[
H'=a\Lambda(I\cdot k)+b(I\cdot S)+c(I\cdot k)(S\cdot k)\] was also
included in the Hamiltonian to reproduce the relative intensities
of the P, Q, and R branches, in agreement with that observed for
Br-HF, although here we do not resolve the individual hyperfine
transitions. The $(I\cdot k)$ term represents the coupling of the
nuclear spin with the axial magnetic field of the molecule, which
is proportional to $\Lambda$, $(I\cdot S)$ is the direct coupling
of the nuclear and electronic spin angular momenta, and $(I\cdot
k)(S\cdot k)$ is a second order interaction of the nuclear and
electronic spins due to both of their projections on the molecular
axis. Assuming Hund's case (a), the expression for the interaction
energy is:

\[
W=[a\Lambda+(b+c)\Sigma]\frac{\Omega}{J(J+1)}I\cdot J\]
 where

\[
I\cdot J=\frac{F(F+1)-J(J+1)-I(I+1)}{2}\] Note that the quantum
number representing nuclear spin (I) equals 3/2 for bromine atoms,
and F (the quantum number representing the total angular momentum
including nuclear spin) is the coupling of J and I. The relative
transition intensities were calculated using spherical tensor matrix
elements similar to those already reported for the Stark effect for
X-HF \cite{14522,14415}, and a Boltzmann distribution for the
rotational state population at 0.37 K. Individual transitions are
convoluted with a 0.024 cm$^{-1}$ Lorentzian lineshape function
which was also varied to obtain the best fit. The spectroscopic
constants ($\nu$, B, D, and {[}a$\Lambda$+ (b+c)$\Sigma$]) resulting
from the fit are summarized in Table \ref{table:X-HCN-experimental}.

\begin{table}
\begin{tabular}{cccc} \hline
Constant& Br-HCN& I-HCN\tabularnewline
\hline \hline $\nu_{0}$ (cm$^{-1}$)& 3284.61(1)&
3277.79(1)\tabularnewline B (cm$^{-1}$)& 0.0151(5)&
0.0120(5)\tabularnewline D (cm$^{-1}$)& 1.5 $\times$ 10$^{-4}$& 1.0
$\times$ 10$^{-4}$\tabularnewline {[}a$\Lambda$ + (b+c)$\Sigma$]
(cm$^{-1}$)& 0.04(1)& 0.04(1)\tabularnewline \hline
\end{tabular}

\caption{\label{table:X-HCN-experimental}A summary of the
experimental parameters for the linear hydrogen bound X-HCN
complexes obtained from a fit to the field-free spectra.}
\end{table}

The experimental (helium) rotational constant is a factor of 2.7
smaller than that predicted from our \emph{ab initio} calculations
due to the fact that some of the helium follows the rotational
motion of the molecule, thus adding to the complexes moment of
inertial \cite{15194}. This magnitude of reduction is in good
agreement with that observed for many other systems. The fact that
the rotational constant of Br-HCN is reduced by a factor of 2.7
while the rotational constant of Br-HF is reduced by a factor of
2.2 is also in agreement that the rotational constant reduction
factor is usually proportional to the magnitude of the B value in
this range \cite{15194}. For Cl-HF, the experimental vibrational
frequencies have provided a very precise benchmark for theoretical
calculations \cite{15189,15190}, and it will be interesting to
compare similar calculations for X-HCN \cite{15248}. The scaled
harmonic vibrational frequency (RMP2/aug-cc-pVTZ) for this isomer
is 3281.78 cm$^{-1}$, in good agreement with that observed
(3284.61 cm$^{-1}$).

The vibrational frequencies of helium solvated rotors are
typically very close to the gas phase values, owing to the weak
interactions of the solvent, thus in general allowing a direct
comparison with theory. A systematic increase in the vibrational
red-shifts of complexes exhibiting linear hydrogen bonds has been
observed however \cite{15194}. While a fully quantitative theory
for this observation is still lacking, such an effect can be
qualitatively understood by the fact that the helium will act to
reduce the amplitude of vibrational averaging, therefore making
the hydrogen bond more linear, and thus stronger, leading to a
greater frequency shift. A second contribution to the observed
red-shift is also expected to arise from a polarization effect due
to the transition dipole moment of HCN, giving rise to dipole -
induced dipole interactions with the solvent, which further lowers
the energy of excited state with respect to the ground state. By
fitting the enhancement of the frequency shift (helium - gas phase
origin (Y)) versus the absolute magnitude of the shift observed in
helium (complex - monomer origin (X)) for a database of 13
complexes observed both in helium and the gas phase, we have
developed an empirical correction factor for the influence of the
helium on the ``gas-phase'' band origin (in cm$^{-1}$):
\begin{equation} \label{eq:origin} Y = 1.822+ 0.03655X
\end{equation} The standard deviation of Y is 1.2 cm$^{-1}$.
See Figure 21 of reference \cite{15194} for a plot of the
experimental data points.
 Correcting for the helium induced shift one arrives at a
predicted gas-phase origin of 3287.40 cm$^{-1}$. Although this
makes the agreement with the scaled harmonic frequency somewhat
worse, it is probably within the error of the \emph{ab initio}
calculations, and we await an estimate for the vibrational
frequency from bound state calculations.

\subsection{I-HCN}

\begin{figure}
\includegraphics{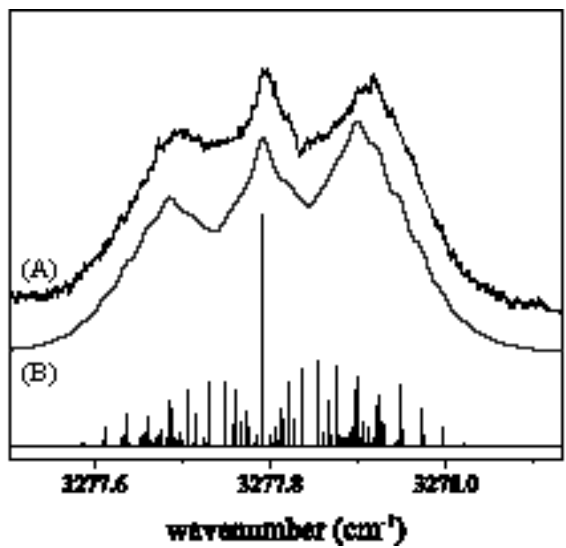}

\caption{\label{fig:I-HCN} (A) A field-free spectrum of the I-HCN
complex in helium droplets. The fit to the spectrum (B), which
includes the effects of nuclear hyperfine (I=5/2 for iodine), was
used to derive the spectroscopic constants given in Table
\ref{table:X-HCN-experimental}. A Lorentzian linewidth of 0.03
cm$^{-1}$ was used for the simulation. }
\end{figure}

A similar study was carried out for the I-HCN complex, the infrared
field-free spectrum being shown in Figure \ref{fig:I-HCN}(A).
 The fit to the spectrum includes the hyperfine effects discussed
above, however, I = 5/2 for iodine. The molecular constants
derived from the fit are given in Table
\ref{table:X-HCN-experimental}. Here again the experimental
rotational constant is consistent with the \emph{ab initio}
calculations, corresponding to a ratio of 2.83. The frequency
shift from HCN monomer is measured to be -33.41 cm$^{-1}$
(estimated gas-phase value using Equation \ref{eq:origin}: -30.37
cm$^{-1}$), in excellent agreement with the calculations employing
the ECP (-31.94 cm$^{-1}$). The nuclear magnetic hyperfine
constant used in the fit is the same as for Br-HCN, namely 0.04
cm$^{-1}$. This is reasonable given that the magnetic moments for
the atoms are quite similar \cite{14319,14320}.

From the survey scans of Figure
\ref{fig:X-X2-HCN-pendular-survey}, peaks corresponding to the
analogous Cl-HCN and F-HCN complexes are not observed. This is in
contrast to the free stretch region where a new peak is observed
for the HCN + Cl case, which will be discussed below.

\section{The HCN-B\lowercase{r} and HCN-I complexes\label{sec:HCN-Xcomplexes}}

\subsection{HCN-Br}

We now turn our attention to the second set of new pendular peaks
that grow in under the appropriate conditions for halogen atom
pickup, namely those that exhibit a much smaller red-shift from
HCN monomer. In the HCN + Br pendular survey scan (Figure
\ref{fig:X-X2-HCN-pendular-survey}), a peak is observed at 3309.55
cm$^{-1}$ which is red-shifted from HCN monomer by -1.65
cm$^{-1}$, in qualitative agreement with the calculations for a
nitrogen bound HCN-Br isomer (-3.14 cm$^{-1}$ at the
RMP2/aug-cc-pVTZ level). Because the ground electronic state of
this complex is predicted to be $^{2}\Sigma_{1/2}$, spin-orbit
coupling could be important in this system, and thus harmonic
frequency calculations (which are performed on the $^{2}\Sigma$
surface) are only semi-quantitative. From our 1D PES's which
include spin-orbit coupling, we find that the binding energy of
the complex is increased by 1.61 cm$^{-1}$ upon ``vibrational
excitation'' of the HCN. Neglecting dynamical coupling of the
different dimensions of the PES, this increase in the binding
energy upon excitation necessitates a red-shift in the vibrational
frequency of equal value, in excellent agreement with that
observed, namely -1.65 cm$^{-1}$. Such excellent agreement could
be fortuitous, and it will be interesting to compare this estimate
with the results from higher dimensional calculations. An
analogous comparison between the $v_{CH}$=0 and $v_{CH}$=1 bound
states for the non-relativistic curves gives a frequency shift of
-2.66 cm$^{-1}$, which is in much poorer agreement with
experiment.

\begin{figure}
\includegraphics{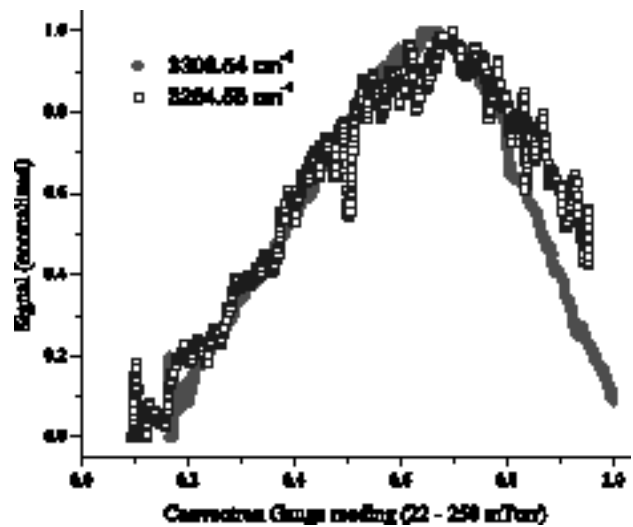}
\caption{\label{fig:HCN-Br-PUC}Br-HCN and HCN-Br pyrolysis source
pressure dependence measurements which aid in the determination that
both observed bands are due to 1:1 complexes.}
\end{figure}

The pick-up of dopants by helium droplets is a statistical process,
and the signal intensity as a function of dopant pressure, a Pick-Up
Cell (PUC) pressure dependence curve, is given by a poisson
distribution weighted by the droplet size distribution for a given
set of source conditions \cite{15194}. PUC curves are a valuable
characterization tool because higher order clusters can easily be
distinguished by their optimum pressure. For our pyrolysis
measurements, the pick-up zone is rather ill defined, however the
optimum molecular halogen pressure upstream of the pyrolysis region
could be used to aid in the determination of this new band to a
second isomer of bromine atoms complexed with HCN. A pressure gauge
was added to the pyrolysis source and the resulting pressure
dependence curves for the two peaks at 3309.54 and 3284.55 cm$^{-1}$
are shown in Figure \ref{fig:HCN-Br-PUC}.
Both bands are found to have exactly the same bromine pressure
dependence, confirming that this new band at 3309.54 cm$^{-1}$ is
also due to a 1:1 complex.

\begin{figure*}
\includegraphics[width=6in]{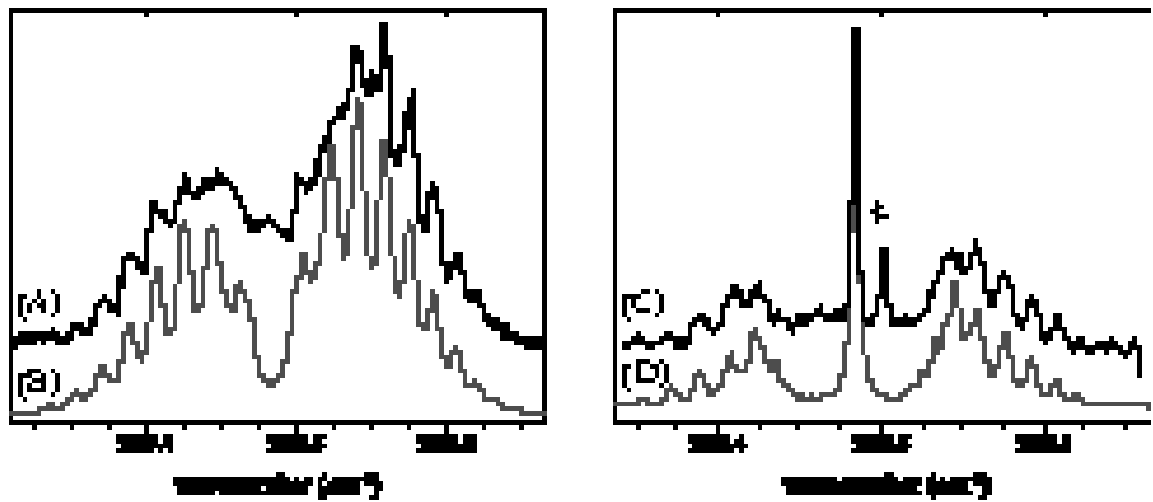}

\caption{\label{fig:HCN-Br-withSOfit} (Left panel) (A) An infrared
field-free spectrum for the CH stretching vibration of the HCN-Br
complex. Fine structure is observed which we preliminarily assign to
a parity splitting induced by spin-orbit coupling of the
$^{2}\Sigma_{1/2}$ and $^{2}\Pi_{1/2}$ states of the complex. A
simulation of the band is also shown (B) which includes the effects
of spin-orbit coupling. (Right panel) (C) The corresponding Stark
spectrum for the HCN-Br complex recorded at an electric field
strength of 2.25 kV cm$^{-1}$. The peak marked with an asterix is
assumed to be related to an impurity complex. Lorentzian linewidths
of 0.025 and 0.01 cm$^{-1}$ were used for fitting the field free and
Stark spectra respectively.}
\end{figure*}

The field-free and Stark spectra for this band are shown in Figure
\ref{fig:HCN-Br-withSOfit}(A) and (C). The band shape is consistent
with that of a linear rotor in a $^{2}\Sigma$ electronic state, as
predicted by our earlier electrostatic arguments.
 In the field-free spectrum there appears to be fine structure, most
notably as a splitting in the low rotational transitions. We
tentatively ascribe this splitting to an interaction of the
(relativistic) $^{2}\Sigma_{1/2}$ and $^{2}\Pi_{1/2}$ states, which
are coupled by the off-diagonal spin-orbit perturbation. The
simulation shown in Figure \ref{fig:HCN-Br-withSOfit} was generated
using an effective Hamiltonian which includes the off-diagonal
spin-orbit interaction derived from one-dimensional calculations,
however the splitting is not readily visible on this scale
\cite{15250}.  Based on preliminary calculations of the bound states
on the 3D spin-orbit corrected diabatic potential energy surfaces,
it appears that the magnitude of the parity splittings for HCN-Br
are smaller in a helium droplet than in the predicted gas-phase
spectrum \cite{15248}.  Interestingly this behavior would be
opposite to that observed recently for NO, where an increase in the
parity splitting was observed \cite{15187}. The spectroscopic
constants derived from the fit to the HCN-Br spectrum are collected
in Table \ref{table:HCN-X-experimental-table}. The rotational
constant obtained from the simulation is 0.019 cm$^{-1}$, which can
be compared with the zero-point corrected (B$_{0}$) values derived
from the bound states on both the non-relativistic ($^{2}\Sigma$)
and relativistic ($^{2}\Sigma_{1/2}$) 1D potential energy surfaces,
namely 0.0623 and 0.0538 cm$^{-1}$ respectively. As shown in the 1D
potentials, and reflected in the B values, the spin-orbit coupling
shifts the potential minimum to slightly longer internuclear
distance. Since a fully quantitative theory for the reduction of
rotational constants due to the helium is still lacking, we cannot
definitively say which of these calculated B values is in better
agreement with experiment, since both reduction factors are within
the range typically observed. By comparing the B reductions for the
HCN-Cl, HCN-Br, and HCN-I complexes however we do gain some insight
into the accuracy of our relativistic potentials; see details below.

A Stark spectrum for HCN-Br is shown in Figure
\ref{fig:HCN-Br-withSOfit}(C), recorded at an applied electric field
strength of 2.25 kV cm$^{-1}$. The simulation includes the effects
of the \emph{e/f} symmetry splitting, however due to the mixing of J
levels, its incorporation only slightly modifies the predicted
spectrum. The peak marked with an asterix is ascribed to an impurity
complex and likely contributes slightly to the congestion in the
field-free spectrum. The ground and excited state dipole moments
obtained from the simulation are 3.78 D, in excellent agreement with
that found at the RMP2/aug-cc-pVTZ level (3.77 D).

\subsection{HCN-I}

\begin{figure*}
\includegraphics[width=6in]{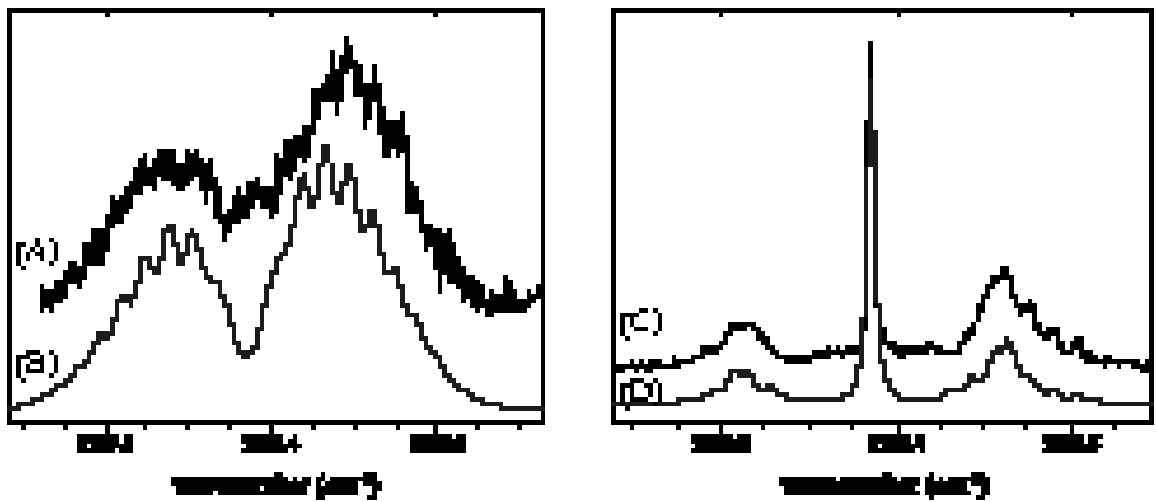}

\caption{\label{fig:HCN-I} The field-free (A) and Stark (C) infrared
spectra of the CH stretching vibration of the HCN-I complex observed
in helium droplets. The simulation of the field free spectrum (B)
was used to determine the molecular constants and includes an
off-diagonal spin-orbit perturbation, which however only slightly
modifies the band shape at this resolution. Lorentzian linewidths of
0.025 and 0.01 cm$^{-1}$ were used for fitting the field free and
Stark spectra respectively.}
\end{figure*}

Based on the pendular survey scans shown in Figure
\ref{fig:X-X2-HCN-pendular-survey}, we observe a pendular peak at
3309.37 cm$^{-1}$ , which we tentatively assign a free stretch
complex between HCN and an iodine atom. The field-free spectrum for
this band is shown in Figure \ref{fig:HCN-I}(A). Again the band
shape is consistent with a molecule with no net orbital angular
momentum (a $\Sigma$ state), in agreement with the results for the
HCN-Br complex.
 The observed frequency shift from HCN monomer is -1.83 cm$^{-1}$,
which agrees nicely with the predicted increase in the binding
energy upon vibrational excitation on the relativistic potentials,
namely 1.63 cm$^{-1}$. The rotational constant derived from the
simulation (B) is B = 0.016 cm$^{-1}$ which is consistent with the
results for HCN-Br. The corresponding value of the rotational
constant derived from the lowest bound state of the
$^{2}\Sigma_{1/2}$ potential is B$_{0}$ = 0.04247 cm$^{-1}$. Figure
\ref{fig:HCN-I}(C) also shows a Stark spectrum for HCN-I. The lines
in the Stark spectrum appear to be much narrower than in the
corresponding field-free scan, which we attribute to considerable
unresolved fine structure in field-free scan, which is decoupled by
the field. Such fine structure is most likely caused by the
spin-orbit coupling in this system, in analogy to that observed for
HCN-Br. The electric field strength used to record the Stark
spectrum (C) was not accurately calibrated and we can only say that
the dipole moment is consistent with the \emph{ab initio} value of
3.91 D.

\section{The HCN-C\lowercase{l} complex\label{sec:The-HCN-Cl-complex}}

\begin{figure*}
\includegraphics[width=6in]{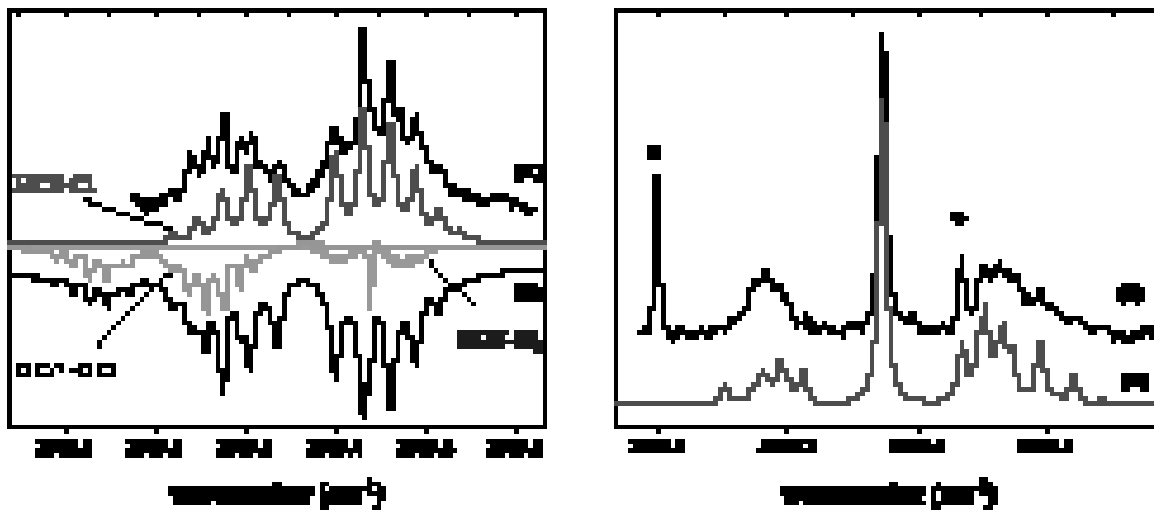}

\caption{\label{fig:HCN-Cl-spec-fit} (Left panel) (A) A rotationally
resolved spectrum of the $\nu_{CH}$ band of the HCN-Cl complex in
helium nanodroplets. The spectrum also has features due to both the
HCN-HCl and HCN-Cl$_{2}$ complexes which overlap in this frequency
region.  Trace (B) is a summation of the 3 bands. (Right panel) (C)
An experimental Stark spectrum of the corresponding band taken in
the presence of a 2.833 kV cm$^{-1}$ electric field. (D) A
simulation of the experimental spectrum based only on the molecular
parameters for HCN-Cl.  The peaks marked with an asterix are due to
the overlapping bands. Lorentzian linewidths of 0.028 and 0.013
cm$^{-1}$ were used for fitting the field free and Stark HCN-Cl
spectra respectively.}
\end{figure*}

When flowing Cl$_{2}$ through the hot pyrolysis source, no bands
were observed that could be assigned to a hydrogen bound Cl-HCN
complex. In the free stretch region however, a new peak is observed
at 3309.33 cm$^{-1}$ in good agreement with the previously observed
bands for the HCN-Br (3309.55 cm$^{-1}$) and HCN-I (3309.37
cm$^{-1}$) complexes. The field-free and Stark spectra for HCN-Cl
are shown in Figure \ref{fig:HCN-Cl-spec-fit}(A) and (C)
respectively. This new peak is only slightly shifted from the
HCN-Cl$_{2}$ band, and thus the pyrolysis source was run at the
highest possible temperatures to reduce the Cl$_{2}$ related
signals. Unfortunately these conditions also lead to an increase in
the amount of HCl produced in the source, presumably from the
reaction of chlorine atoms with the quartz pyrolysis tube, and the
HCN-HCl molecular complex is also observed. The assignment of this
band to HCN-HCl was confirmed by a separate experiment in which HCl
was intentionally introduced into the pick-up cell.
 Despite the overlap of two impurity complexes, we are able to reproduce
all of the features of the spectrum, and extract the molecular
parameters which are listed in Table
\ref{table:HCN-X-experimental-table}. In the analysis of the
spectrum we were not able to identify any fine structure that
could be attributed to a spin-orbit interaction, which we
tentatively assigned in the bromine and iodine complexes. This is
consistent with the much smaller spin-orbit coupling constant for
atomic chlorine compared to bromine or iodine. Indeed a simulation
of the HCN-Cl spectra using the effective Hamiltonian developed
for the HCN-Br, shows no such splittings at the experimental
resolution.

The vibrational band origin of HCN-Cl is shifted from HCN monomer
by -1.87 cm$^{-1}$. The \emph{ab initio} frequency calculations
predict a shift of -2.09 cm$^{-1}$ at the RMP2/aug-cc-pVTZ level
(see Table \ref{table:HCN-X-MP2}) in good agreement with the
experimental value. The shift predicted from the change in the
binding energy upon vibrational excitation from 1D potential
energy surfaces is -2.08 cm$^{-1}$. This value is nearly identical
to the non-relativistic results due to the relatively small
spin-orbit coupling constant for chlorine atoms. Bound state
calculations predict that the isotope splitting between
HCN-$^{35}$Cl and HCN-$^{37}$Cl is approximately 0.001 cm$^{-1}$,
which is negligible at our resolution. In contrast, the two
isotopes of Cl-HF were found to have a much larger splitting
(0.038 cm$^{-1}$), due to the fact that it is a hydrogen bound
complex \cite{14522,15190}.

The rotational constant deduced from the fit to the spectrum is
0.032 cm$^{-1}$. Because the spin-orbit coupling is small in this
case, the predicted rotational constants from the bound states on
the $^{2}\Sigma$ and $^{2}\Sigma_{1/2}$ potentials are very
similar, namely B$_{0}$= 0.08478 (0.08784) cm$^{-1}$ for
HCN-$^{35}$Cl on the $^{2}\Sigma_{1/2}$ ($^{2}\Sigma$) potential.
The reduction of the rotational constant upon solvation is thus a
factor of approximately 2.7. If we assume that the solvation
behavior between the HCN-Cl, HCN-Br, and HCN-I complexes is the
same, then this reduction factor also gives support to our earlier
comparison of HCN-Br and HCN-I with the B values obtained from the
spin-orbit corrected potentials. On these potentials the reduction
factor was also determined to be 2.75, however if we had compared
the experimental value to the B value from the non-relativistic
potentials, the B reduction would have been 3.2.

Figure \ref{fig:HCN-Cl-spec-fit}(C) also shows a Stark spectrum for
HCN-Cl recorded at an applied field strength of 2.833 kV cm$^{-1}$.
The dipole moment determined from the fit is 3.0(2) D, which is
significantly smaller than the \emph{ab initio} value of 3.66 D. It
is interesting to note that the \emph{ab initio} calculations are in
better agreement for bromine than for chlorine what could be due to
vibrational averaging over the low frequency bending and stretching
modes of the complex. This would be more important for Cl due to its
lighter mass,  reducing the dipole moment from its equilibrium
value. Indeed a similar trend was also observed for the X-HF
complexes.

\begin{table}
\begin{tabular}{cccc} \hline Constant& HCN-Cl& HCN-Br&
HCN-I\tabularnewline \hline \hline $\nu$ (cm$^{-1}$)& 3309.33&
3309.55& 3309.37\tabularnewline B (cm$^{-1}$)& 0.032& 0.019&
0.016\tabularnewline D (cm$^{-1}$)& 5.0 $\times$ 10$^{-5}$& 1.2
$\times$ 10$^{-5}$& 1.0 $\times$ 10$^{-5}$\tabularnewline $\mu$ (D)&
3.0& 3.78& -\tabularnewline \hline
\end{tabular}

\caption{\label{table:HCN-X-experimental-table}The experimental
spectroscopic constants obtained from a fit to the infrared
spectra for each of the pre-reactive HCN-X complexes. Fine
structure is observed in the HCN-Br and HCN-I spectra, which is
attributed to spin-orbit interaction. A more detailed theoretical
treatment for this effect is given elsewhere \cite{15250}.}
\end{table}

\section{Di-radical B\lowercase{r}-HCCCN-B\lowercase{r} complexes}

Given the observation of the HCN-Br, Br-HCN, HCN-I, and I-HCN
complexes, it is interesting to consider the possibility of picking
up two halogen atoms, to form a di-radical complex such as
Br-HCN-Br, which utilizes the HCN as a stabilizer, preventing the
recombination to form HCN + Br$_{2}$.  Such complexes would
represent the beginnings of nano-scale radical solids which could be
used as high-energy density materials \cite{13724}. Gordon \emph{et
al.} proposed a similar mechanism was responsible for the
stabilization of nitrogen atoms in a N$_{2}$ matrix (of up to 10\%
atomic concentration) frozen in bulk liquid helium
\cite{6097,10884,10833,10832,10816,10022}. While we have yet to
perform extensive searches for HCN, we did explore this possibility
for Br atoms and cyanoacetylene (HCCCN), where better signals are
observed.
\begin{figure*}
\includegraphics[width=5in]{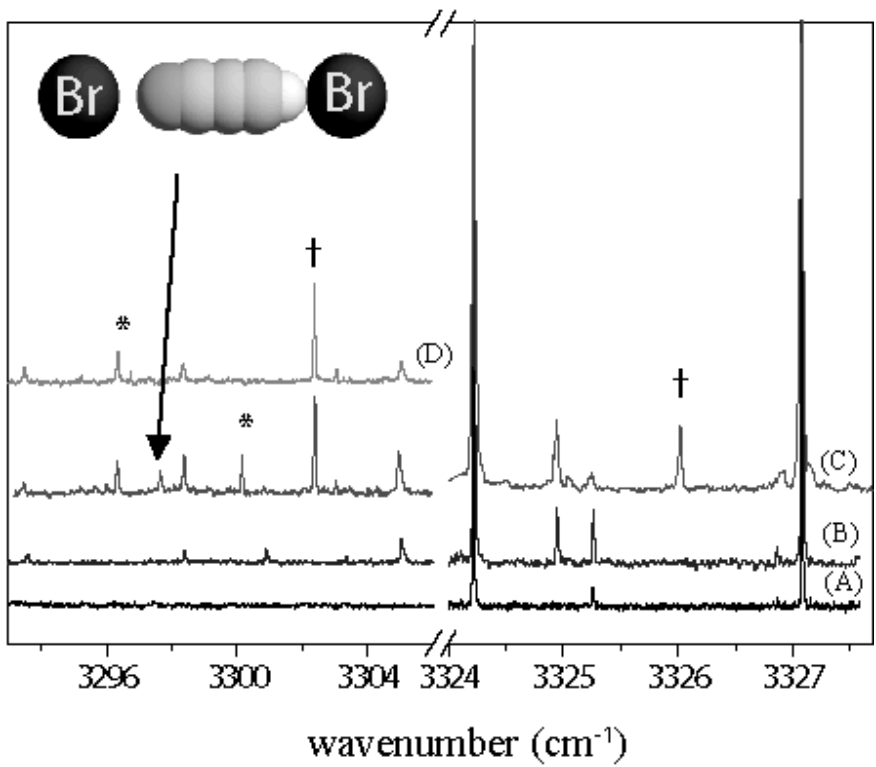}

\caption{\label{fig:Br-HCCCN-Br} A series of pendular scans for Br +
HCCCN where the bromine pressure is intentionally high to facilitate
forming larger clusters. In (A) only HCCCN is picked up by the
droplets and in (B) Br$_{2}$ is flowing through the cold pyrolysis
source. Scans (C) and (D) were recorded with the pyrolysis source at
the appropriate temperature for bromine atom production, but for (D)
the bromine is picked up first where as in (C) the cyanoacetylene is
picked up first. The two peaks marked with a $\dagger$ are assigned
to the linear Br-HCCCN and HCCCN-Br complexes, respectively. The two
peaks marked with an {*} are found to optimize at higher HCCCN
pressure and therefore correspond to more than one HCCCN. The peak
at 3297.65 cm$^{-1}$ optimizes at the same HCCCN pressure as the 1:1
complexes however is found to optimize at higher bromine pressure,
suggesting that it is a complex containing two bromine radicals. The
different behavior based on pick-up order is expected, since in (D)
two bromine atoms will likely recombine to form Br$_{2}$ in the
absence of the HCCCN.}
\end{figure*}
Figure \ref{fig:Br-HCCCN-Br} shows a set of pendular survey scans
covering the important frequency region. It is important to note
that scans \ref{fig:Br-HCCCN-Br}(A) - (C) were recorded by picking
up the HCCCN first, but in \ref{fig:Br-HCCCN-Br}(D) the order has
been reversed, and we instead pickup from the pyrolysis source
first. In Figure \ref{fig:Br-HCCCN-Br}(A), only HCCCN is added to
the droplets, while in \ref{fig:Br-HCCCN-Br}(B), Br$_{2}$ is flowing
through a room temperature pyrolysis source. The result of heating
the pyrolysis to the appropriate temperatures for bromine atom
pick-up are shown in Figures \ref{fig:Br-HCCCN-Br}(C) and (D). In
good agreement with the results of X + HCN, we find two peaks at
3326.1 and 3302.4 cm$^{-1}$ (labeled by a $\dagger$ in the figure)
which we assign to the Br-HCCCN and HCCCN-Br complexes, based on
their frequency shifts and signal strengths. The peaks labeled with
an {*} were found to optimize at higher HCCCN pressure, so they
correspond to complexes containing more than one HCCCN. The peak at
3297.65 cm$^{-1}$ is a good candidate for being Br-HCCCN-Br because
it only appears in the spectrum when the HCCCN is picked up first.
This dependence on pick-up order is to be expected because two
bromine atoms will likely recombine to form Br$_{2}$ in the absence
of the molecular spacer. To aid in this preliminary assignment we
performed bromine pressure dependence measurements which are shown
in Figure \ref{fig:Br-HCCCN-Br-PUC}.

\begin{figure}
\includegraphics{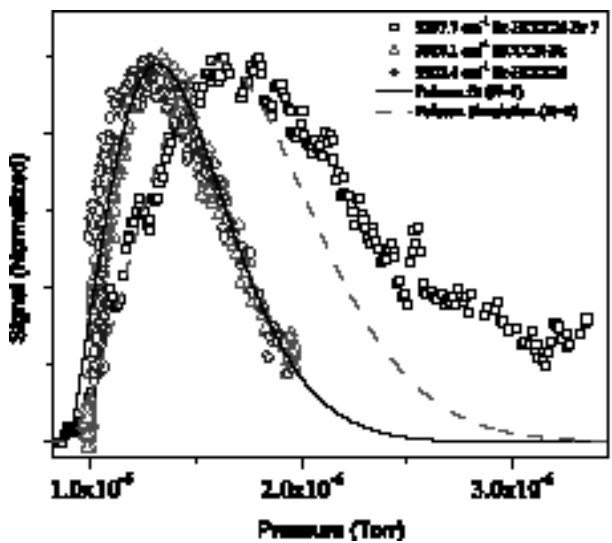}

\caption{\label{fig:Br-HCCCN-Br-PUC} Bromine pressure dependencies
for the peaks assigned to the Br-HCCCN, HCCCN-Br, and Br-HCCCN-Br
complexes. The fact that the peak at 3297.65 cm$^{-1}$ optimizes at
higher bromine pressure than the 1:1 complexes lends supporting
evidence to our claim that this is a di-radical complex. The smooth
curves are the calculated pick-up probabilities obtained by fitting
a Poisson distribution to the 1:1 curve, and then re-simulating the
curve for the dimer. }
\end{figure}
This new peak is found to optimize at higher bromine pressures than
those peaks assigned to the 1:1 complexes, and agrees well with a
simulation of a dimer using a Poisson distribution for the pick-up
statistics. For these pick-up ``cell'' pressure dependence
measurements the main chamber ion-gauge was used as the
\emph{x}-axis, however since this really monitors recombined
Br$_{2}$ after collisions with the walls, instead of the local
pressure of Br atoms in the pick-up zone, the pick-up curve for the
dimer does not correspond to twice the optimum monomer pressure.
Although we cannot rule out that a Br-HCCCN-Br$_{2}$, or similar
complex might have such a pick-up cell pressure dependence, a
double-resonance population transfer experiment \cite{14556} in
which this new peak is pumped and the HCCCN-Br$_{2}$ complex is
recovered would be definitive.

\section{Discussion}

Extensive pendular survey scans have been performed in order to
find evidence for the entrance channel complexes of fluorine atoms
with HCN. From the \emph{ab initio} calculations presented above,
it is clear that the reaction to form the HFCN product is quite
exothermic, and exhibits only a small barrier, if any, to the
reaction. The calculated energies of the HFCN product and its
barrier relative to F + HCN are -34.05 and -2.45 kcal mol$^{-1}$
(-28.44 and 3.05 kcal mol$^{-1}$), respectively, at the G2
(UCCSD(T)/6-311++G(d,p)) level of theory. Two entrance channel
complexes (F-HCN and HCN-F) were also found in the G2 calculations
with energies -0.55 and -1.58 kcal mol$^{-1}$ respectively. The
unintuitive result that the transition state is calculated (at the
G2 level) to be lower in energy than the corresponding van der
Waals minima is simply an artifact of the calculation, and this
small energy difference is within the estimated accuracy of the
method. The fact that we do not observe a fluorine atom
pre-reactive complex could be an indication that indeed the
insertion reaction takes place, even under our experimental
conditions at 0.37 K. This conclusion is also partially supported
by the argon matrix work of Andrews \cite{15155} and Misochko
\cite{15156} where the insertion product has been identified. No
pre-reactive complexes were observed in these works. Guided by
their observed frequency of the CH stretching vibration of 3018
cm$^{-1}$, we searched for HFCN in helium. No signals were
observed, however this frequency range is at the edge of the
tuning range of the lasers used in this work. The oscillator
strength for this vibration is also approximately 50 times weaker
than the HCN fundamental, based on harmonic frequency
calculations, and thus the argon matrix experiment has the big
advantage of concentrating the product using longer deposition
times. We hope to revisit this experiment using newer technology
OPO lasers, with a larger tuning range and higher output power
\cite{15257,14008}.

\begin{figure*}
\includegraphics[width=4in,angle=-90]{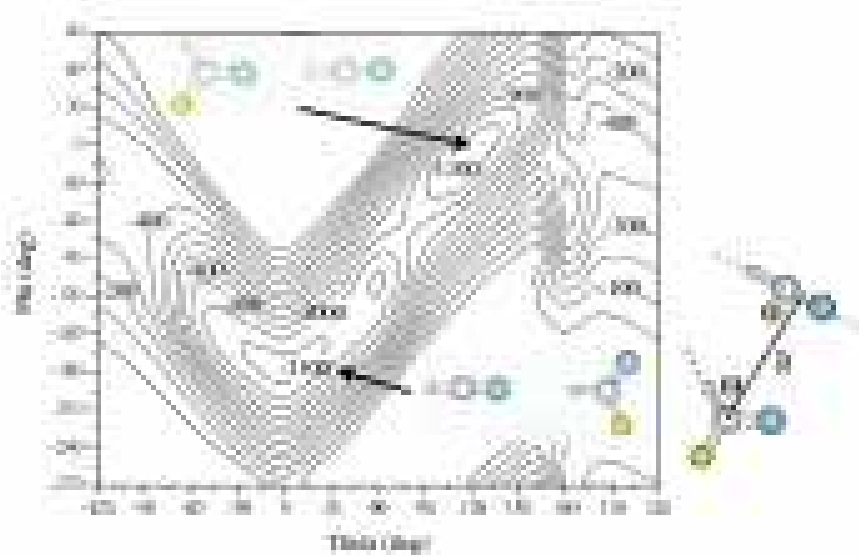}

\caption{\label{fig:HFCN-HCN-PES}A two-dimensional interaction
potential calculated at the RMP2/aug-cc-pVDZ level for the insertion
product HFCN with a second HCN molecule. The second HCN molecule was
assumed to lie in the plane of HFCN due to the fact that no
non-planar geometries were observed in fully relaxed optimizations.
The two angles were stepped in increments of 15\textdegree \ and
only the intermolecular distance was allowed to optimize in the
calculation. Counterpoise correction was applied and a spline
interpolation used to generate the final surface. Contour lines are
drawn in 100 cm$^{-1}$ intervals. By tagging the F + HCN reaction
product with a second HCN molecule, we regain a strong
high-frequency CH stretch which could be then probed
experimentally.}
\end{figure*}

Due to the statistical nature of the pick-up process it should be
possible to ``tag'' the HFCN product with another weakly bound
molecule, which could then act as a stronger infrared chromophore.
Indeed we have preliminary evidence that such a situation also
arises in reaction of aluminum atoms with HCN and HF \cite{15249}.
Shown in Figure \ref{fig:HFCN-HCN-PES} is a two-dimensional
angular potential for HCN + HFCN calculated at the
RMP2/aug-cc-pVDZ level, which shows that multiple van der Waals
isomers may indeed exist. Since cluster formation in helium occurs
sequentially and obeys Poisson statistics, these clusters should
be formed when picking up HCN after the fluorine atom, given that
the droplet is large enough to stabilize the HFCN. Based on the
calculated exothermicity, the formation of HFCN would boil off
approximately 2500 helium atoms (assuming 5 cm$^{-1}$ per atom
\cite{5665}).

The last remaining question in this study is the existence of a
hydrogen bound Cl-HCN complex in analogy with that observed for
bromine and iodine atoms. Due to the fact that the isomerization
barriers between the hydrogen and nitrogen bound isomers are so
small, one possible explanation is that this isomer is not formed
due to a zero-point energy effect, and that the complex simply
converts to the global minimum HCN-Cl isomer. To make any further
analysis of this possibility we must await the spin-orbit
corrected 2D potentials \cite{15248}, to get a more realistic view
of how the isomerization barrier is effected by spin-orbit
coupling. A second possibility is that the reactivity is enhanced
when the chlorine atom approaches the hydrogen end of HCN and it
reacts, in analogy with that proposed for F + HCN.

\section{Summary}

We have presented here a combined experimental and theoretical
investigation of the halogen atom - HCN entrance channel complexes
in order to compare with the X-HF complexes studied previously.
The potential energy surfaces for X + HCN are found to be
considerably more complex than for X+HX, due to the possibility of
hydrogen atom abstraction and addition reactions. The lone pair on
the nitrogen of HCN is also found to significantly alter the shape
of the long-range potential energy surfaces, and for X + HCN, we
find that the global minimum is a linear nitrogen bound geometry
in contrast to the hydrogen bound global minimum for X-HF. A
linear hydrogen bound X-HCN complex is also predicted to be
stable, however the isomerization barrier back to the global
minimum is small. \emph{Ab initio} calculations predict that for
iodine and bromine significant barriers to the chemical reactions
exist, and thus the entrance channel complexes should be quite
stable when formed in a helium droplet. Both of the linear van der
Waals isomers predicted by theory are observed experimentally for
bromine and iodine atoms. Interestingly a HCN-Cl isomer was also
observed whereas Cl-HCN was not. Since HCN-Cl is predicted to be
the global minimum on the van der Waals potential, it could be
that the isomerization barrier is simply too small to prevent
Cl-HCN from rearranging to HCN-Cl. A second possibility is that
the reactivity is enhanced on the hydrogen end of HCN, and Cl-HCN
goes on to react to form HClCN. No entrance channel complexes were
observed in the fluorine experiments, and we preliminarily
interpret this as a result of reaction. The intermediate reaction
product HFCN has been recently observed in an argon matrix,
however, also in that work no pre-reactive complexes were
observed. Preliminary searches for HFCN in this work in helium
nanodroplets did not reveal this species. The photon energy used
to study the CH stretching vibration of the entrance channel
complexes is greater than the predicted barriers, suggesting that
photo-initiation of the corresponding reactions could be possible.
The unique asymptotic degeneracy of the multiple electronic states
in the region of the entrance and exit channels for free radical -
molecule complexes is found to give rise to fine structure in the
observed spectra for HCN-Br and HCN-I.

\section{Acknowledgments} The authors wish to thank Anna Fishchuk and
Ad van der Avoird for their theoretical work on these X-HCN systems
and its correspondence before publication. This work was supported by
the Air Force Office of Scientific Research (AFOSR). Partial support
is also acknowledged from the National Science Foundation
(CHE-99-87740). J.\ K.\ acknowledges financial support of the
Alexander von Humboldt foundation through a Feodor Lynen fellowship.

\bibliography{miller}

\begin{thebibliography}{10}

\bibitem{110}
J.~M. Hutson, {\em Annu. Rev. Phys. Chem.}, 1990, {\bf 41}, 123--154.

\bibitem{13487}
M.~I. Lester, B.~V. Pond, M.~D. Marshall, D.~T. Anderson, L.~B. Harding, and
  A.~F. Wagner, {\em Faraday Disc.}, 2001, {\bf 118}, 373--385.

\bibitem{10758}
Y.~L. Chen and M.~C. Heaven, {\em J. Chem. Phys.}, 1998, {\bf 109}(13),
  5171--5174.

\bibitem{15164}
M.~C. Heaven, {\em Int. Rev. Phys. Chem.}, 2005, {\bf 24}, 375--420.

\bibitem{14697}
T.~C. McInnis and L.~Andrews, {\em J. Phys. Chem.}, 1992, {\bf 96}(5),
  2051--2059.

\bibitem{14484}
E.~Y. Misochko, V.~A. Benderskii, A.~U. Goldschleger, A.~V. Akimov, and A.~F.
  Shestakov, {\em J. Am. Chem. Soc.}, 1995, {\bf 117}, 11997--11998.

\bibitem{14009}
R.~D. Hunt and L.~Andrews, {\em J. Chem. Phys.}, 1988, {\bf 88}, 3599--3606.

\bibitem{10818}
K.~Nauta and R.~E. Miller, {\em Science}, 1999, {\bf 283}, 1895--1897.

\bibitem{11603}
K.~Nauta and R.~E. Miller, {\em Science}, 2000, {\bf 287}, 293--295.

\bibitem{14433}
F.~Madeja, M.~Havenith, K.~Nauta, and R.~E. Miller, {\em J. Chem. Phys.}, 2004,
  {\bf 120}, 10554.

\bibitem{5971}
F.~Stienkemeier, W.~E. Ernst, J.~Higgins, and G.~Scoles, {\em J. Chem. Phys.},
  1995, {\bf 102}, 615--617.

\bibitem{14522}
J.~M. Merritt, J.~K\"upper, and R.~E. Miller, {\em Phys. Chem. Chem. Phys.},
  2005, {\bf 7}(1), 67--78.

\bibitem{14912}
J.~M. Merritt, S.~Rudi\'c, and R.~E. Miller, {\em J. Chem. Phys.}, 2006, {\bf
  124}, 084301--084313.

\bibitem{15251}
S.~Rudi\'c, J.~M. Merritt, and R.~E. Miller, {\em J. Chem. Phys.}, 2006, {\bf
  124}, 104305.

\bibitem{14514}
N.~Balakrishnan, {\em J. Chem. Phys.}, 2004, {\bf 121}(12), 5563--5566.

\bibitem{14281}
D.~Skouteris, D.~E. Manolopoulos, W.~Bian, H.~J. Werner, L.~H. Lai, and K.~Liu,
  {\em Science}, 1999, {\bf 286}, 1713--1716.

\bibitem{15252}
P.~F. Weck and N.~Balakrishnan, {\em Int. Rev. Phys. Chem.}, 2006, {\bf 25},
  283.

\bibitem{14832}
D.~Towsend, S.~A. Lahankar, S.~K. Lee, S.~D. Chambreau, A.~G. Suits, X.~Zhang,
  J.~Rheinecker, L.~B. Harding, and J.~M. Bowman, {\em Science}, 2005, {\bf
  306}, 1158.

\bibitem{15239}
X.~Zhang, J.~Rheinecker, and J.~M. Bowman, {\em J. Chem. Phys.}, 2005, {\bf
  122}, 114313.

\bibitem{9385}
C.~S. Maierle, G.~C. Schatz, M.~S. Gordon, P.~Mccabe, and J.~N. Connor, {\em J.
  Chem. Soc. Faraday Trans.}, 1997, {\bf 93}, 709--720.

\bibitem{14015}
W.~B. Zeimen, J.~Klos, G.~C. Groenenboom, and A.~van~der Avoird, {\em J. Phys.
  Chem. A}, 2003, {\bf 107}, 5110--5121.

\bibitem{15253}
G.~W.~M. Vissers and A.~B. McCoy, {\em J. Phys. Chem. A}, 2006, {\bf 110},
  5978--5981.

\bibitem{15256}
W.~B. Zeimen, J.~Klos, G.~C. Groenenboom, and A.~van~der Avoird, {\em J. Phys.
  Chem. A}, 2004, {\bf 107}, 9319--9322.

\bibitem{15254}
M.~P. Deskevich, M.~Y. Hayes, K.~Takahashi, R.~T. Skodje, and D.~J. Nesbitt,
  {\em J. Chem. Phys.}, 2006, {\bf 124}, 224303.

\bibitem{13847}
P.~Zdanska, D.~Nachtigallova, P.~Nachtigall, and P.~Jungwirth, {\em J. Chem.
  Phys.}, 2001, {\bf 115}(13), 5974--5983.

\bibitem{11590}
M.~Meuwly and J.~M. Hutson, {\em J. Chem. Phys.}, 2000, {\bf 112}(2), 592--600.

\bibitem{15255}
J.~Klos, M.~M. Szczesniak, and G.~Chalasinski, {\em Int. Rev. Phys. Chem.},
  2004, {\bf 23}, 541--571.

\bibitem{12327}
C.~Kreher, R.~Theinl, and K.~H. Gericke, {\em J. Chem. Phys.}, 1996, {\bf
  104}(12), 4481.

\bibitem{13257}
B.~K. Decker, G.~He, I.~Tokue, and R.~G. Macdonald, {\em J. Phys. Chem. A},
  2001, {\bf 105}(24), 5759--5767.

\bibitem{5577}
R.~B. Metz, J.~M. Pfeiffer, J.~D. Thoemke, and F.~F. Crim, {\em Chem. Phys.
  Lett.}, 1994, {\bf 221}, 347--352.

\bibitem{13342}
D.~Troya, M.~Gonzalez, G.~S. Wu, and G.~C. Schatz, {\em J. Phys. Chem. A},
  2001, {\bf 105}(11), 2285--2297.

\bibitem{15240}
L.~B. Harding, {\em J. Phys. Chem.}, 1996, {\bf 100}, 10123--10130.

\bibitem{15241}
J.~de~Juan, S.~Callister, H.~Reisler, G.~A. Segal, and C.~Wittig, {\em J. Chem.
  Phys.}, 1988, {\bf 89}, 1977--1985.

\bibitem{15243}
I.~R. Sims and I.~W.~M. Smith, {\em J. Chem. Soc. Faraday Trans. 2}, 1989, {\bf
  85}, 915--923.

\bibitem{12458}
M.~J. Frost, I.~W.~M. Smith, and R.~D. Spencer-Smith, {\em J. Chem. Soc.
  Faraday Trans. 2}, 1993, {\bf 89}, 2355.

\bibitem{7706}
J.~M. Pfeiffer, R.~B. Metz, J.~D. Thoemke, E.~Woods, and F.~F. Crim, {\em J.
  Chem. Phys.}, 1995, {\bf 104}, 4490--4501.

\bibitem{9220}
C.~Kreher, J.~L. Rinnenthal, and K.~H. Gericke, {\em J. Chem. Phys.}, 1998,
  {\bf 108}, 3154--3167.

\bibitem{15155}
R.~D. Hunt and L.~Andrews, {\em Inorg. Chem.}, 1987, {\bf 26}, 3051--3054.

\bibitem{15156}
I.~U. Goldschleger, A.~V. Akimov, E.~Y. Misochko, and C.~A. Wight, {\em
  Mendeleev Comm.}, 2001, (2), 43--45.

\bibitem{11280}
K.~Nauta and R.~E. Miller, {\em J. Chem. Phys.}, 1999, {\bf 111}, 3426--3433.

\bibitem{10872}
E.~L. Knuth, B.~Schilling, and J.~P. Toennies, Oxford University Press, Oxford,
  1995;
\newblock Vol. ~19, pp. 270--276.

\bibitem{13231}
J.~K\"upper, J.~M. Merritt, and R.~E. Miller, {\em J. Chem. Phys.}, 2002, {\bf
  117}(2), 647--652.

\bibitem{12851}
K.~Nauta and R.~E. Miller, {\em J. Chem. Phys.}, 2001, {\bf 115}(22),
  10138--10145.

\bibitem{5665}
D.~M. Brink and S.~Stringari, {\em Z. Phys. D}, 1990, {\bf 15}, 257--263.

\bibitem{14143}
G.~E. Douberly and R.~E. Miller, {\em J. Phys. Chem. B}, 2003, {\bf 107}(19),
  4500--4507.

\bibitem{4713}
S.~F. Boys and F.~Bernardi, {\em Mol. Phys.}, 1970, {\bf 19(4)}, 553--566.

\bibitem{13608}
A.~C. Legon and J.~C. Thorn, {\em J. Chem. Soc. Faraday Trans.}, 1993, {\bf
  89}(23), 4157--4162.

\bibitem{14561}
K.~Hinds and A.~C. Legon, {\em Chem. Phys. Lett.}, 1995, {\bf 240}, 467--473.

\bibitem{14562}
A.~C. Legon and K.~Hinds, {\em Mol. Phys.}, 1996, {\bf 88}(3), 673--682.

\bibitem{14497}
L.~A. Curtiss, K.~Raghavachari, G.~W. Trucks, and J.~A. Pople, {\em J. Chem.
  Phys.}, 1991, {\bf 94}(11), 7221.

\bibitem{12826}
M.~W. {Chase Jr}, {\em J. Phys. Chem. Ref. Data}, 1998, {\bf 4}(9), 1--1951.

\bibitem{15248}
A.~Fishchuk, J.~M. Merritt, and A.~van~der Avoird, in preparation, 2006.

\bibitem{13953}
H.~J. Werner, P.~J. Knowles, R.~D. Amos, A.~Bernhardsson, A.~Berning,
  P.~Celani, D.~L. Cooper, M.~J.~O. Deegan, A.~J. Dobbyn, F.~Eckert, C.~Hampel,
  G.~Hetzer, T.~Korona, R.~Lindh, A.~W. Lloyd, S.~J. McNicholas, F.~R. Manby,
  W.~Meyer, M.~E. Mura, A.~Nicklass, P.~Palmieri, R.~Pitzer, G.~Rauhut,
  M.~Schutz, U.~Schumann, H.~Stoll, A.~J. Stone, R.~Tarroni, and
  T.~Thorsteinsson, {\em MOLPRO, A package of Ab initio Programs, version
  2002.1}, University College Cardiff Consultants Limited, Wales, UK, 2002.

\bibitem{14016}
J.~Klos, G.~Chalasinski, H.~Werner, and M.~M. Szczesniak, {\em J. Chem. Phys.},
  2003, {\bf 115}(7), 3085--3098.

\bibitem{15193}
F.~M. Tao and Y.~K. Pan, {\em J. Chem. Phys.}, 1992, {\bf 97}(7), 4989--4995.

\bibitem{15189}
A.~V. Fishchuk, P.~E.~S. Wormer, and A.~van~der Avoird, {\em J. Phys. Chem. A},
  2006, {\bf 110}(16), 5273--5279.

\bibitem{14349}
K.~A. Peterson, D.~Figgen, E.~Goll, H.~Stoll, and M.~Dolg, {\em J. Chem.
  Phys.}, 2003, {\bf 119}(21), 11113--11123.

\bibitem{12216}
M.~Meuwly and J.~M. Hutson, {\em Phys. Chem. Chem. Phys.}, 2000, {\bf 2}(4),
  441--446.

\bibitem{1228}
H.~Lefebvre-Brion and R.~W. Field, {\em Perturbations in the Spectra of
  Diatomic Molecules}, Academic Press, Orlando, Fl., 1986.

\bibitem{15195}
J.~E. Rode, J.~Klos, L.~Rajchel, M.~Szczeniak, G.~Chalasinski, and A.~A.
  Buchachenko, {\em J. Phys. Chem. A}, 2005, {\bf 109}, 11484--11494.

\bibitem{15210}
A.~Nicklass, K.~A. Peterson, A.~Berning, H.~J. Werner, and P.~J. Knowles, {\em
  J. Chem. Phys.}, 2000, {\bf 112}(13), 5624--5632.

\bibitem{15165}
R.~J. LeRoy, Level 7.7: A computer program for solving the radial schr\"odinger
  equation for bound and quasibound levels University of Waterloo Chemical
  Physics Research Report CP-661; see the ``computer programs'' link at
  http://leroy.uwaterloo.ca, 2005.

\bibitem{11279}
K.~Nauta, D.~T. Moore, and R.~E. Miller, {\em Faraday Disc.}, 1999, {\bf 113},
  261--278.

\bibitem{13829}
E.~W. Draeger and D.~M. Ceperley, {\em Phys. Rev. Lett.}, 2003, {\bf 90}(6),
  065301.

\bibitem{14306}
R.~A. Frosh and H.~M. Foley, {\em Phys. Rev.}, 1952, {\bf 88}(6), 1337--1349.

\bibitem{14415}
J.~Brown and A.~Carrington, {\em Rotational Spectroscopy of Diatomic
  Molecules}, Cambridge molecular science, 2003.

\bibitem{15194}
M.~Y. Choi, G.~E. Douberly, T.~M. Falconer, W.~K. Lewis, C.~M. Lindsay, J.~M.
  Merritt, P.~L. Stiles, and R.~E. Miller, {\em Int. Rev. Phys. Chem.}, 2006.

\bibitem{15190}
A.~V. Fishchuk, G.~C. Groenenboom, and A.~van~der Avoird, {\em J. Phys. Chem.
  A}, 2006, {\bf 110}(16), 5280--5288.

\bibitem{14319}
V.~Jaccarino, J.~G. King, R.~A. Satten, and H.~H. Stroke, {\em Phys. Rev.},
  1954, {\bf 94}, 1798--1799.

\bibitem{14320}
H.~H. Brown and J.~G. King, {\em Phys. Rev.}, 1966, {\bf 142}(1), 53--59.

\bibitem{15250}
J.~M. Merritt and R.~W. Field, in preparation, 2006.

\bibitem{15187}
K.~von Haeften, A.~Metzelthin, S.~Rudolph, V.~Staemmler, and M.~Havenith, {\em
  Phys. Rev. Lett.}, 2005, {\bf 95}(21), 215301.

\bibitem{13724}
F.~Cacace, G.~de~Petris, and A.~Troiani, {\em Science}, 2002, {\bf 295},
  480--481.

\bibitem{6097}
E.~B. Gordon, V.~V. Khmelenko, A.~A. Pelmenev, E.~A. Popov, and O.~F. Pugachev,
  {\em Chem. Phys. Lett.}, 1989, {\bf 155}, 301--304.

\bibitem{10884}
R.~E. Boltnev, E.~B. Gordon, V.~V. Khmelenko, I.~N. Krushinskaya, M.~V.
  Martynenko, A.~A. Pelmenev, E.~A. Popov, and A.~F. Shestakov, {\em Chem.
  Phys.}, 1994, {\bf 189}, 367--382.

\bibitem{10833}
E.~B. Gordon, A.~A. Pelmenev, O.~F. Pugachev, and V.~V. Khmelenko, {\em Chem.
  Phys.}, 1981, {\bf 61}, 35--41.

\bibitem{10832}
E.~B. Gordon, L.~P. Mezhov-Deglin, O.~F. Pugachev, and V.~V. Khmelenko, {\em
  Chem. Phys. Lett.}, 1978, {\bf 54}(2), 282--285.

\bibitem{10816}
E.~B. Gordon, V.~V. Khmelenko, A.~A. Pelmenev, E.~A. Popov, O.~F. Pugachev, and
  A.~F. Shestakov, {\em Chem. Phys.}, 1993, {\bf 170}, 411--426.

\bibitem{10022}
R.~E. Boltnev, E.~B. Gordon, I.~N. Krushinskaya, M.~V. Martynenko, A.~A.
  Pelmenev, E.~A. Popov, V.~V. Khmelenko, and A.~F. Shestakov, {\em Low Temp.
  Phys.}, 1997, {\bf 23}, 567--577.

\bibitem{14556}
G.~E. Douberly, J.~M. Merritt, and R.~E. Miller, {\em Phys. Chem. Chem. Phys.},
  2005, {\bf 7}(3), 463--468.

\bibitem{15257}
J.~M. Merritt, {\em Ph.D. Dissertation, University of North Carolina at Chapel
  Hill}, 2006.

\bibitem{14008}
F.~J.~M. Harren, S.~Li, S.~E. Bisson, and M.~M. J. W.~V. Herpen, {\em Appl.
  Phys. B}, 2002, {\bf 75}, 329--333.

\bibitem{15249}
J.~M. Merritt, G.~E. Douberly, P.~L. Stiles, and R.~E. Miller, in preparation,
  2006.

\end{thebibliography}

\end{document}